\documentclass[manuscript=article]{achemso}
\usepackage{achemso}
\setkeys{acs}{usetitle = true}
\usepackage[version=3]{mhchem} 
\usepackage[T1]{fontenc}       
\usepackage{color,soul}
\usepackage{amsmath,amssymb}
\usepackage{subfig}
\usepackage{natbib}
\usepackage{multirow}
\DeclareUnicodeCharacter{0308}{~}

\author{Arunima Singh$^{*}$, Manjari Jain, Saswata Bhattacharya} 
\affiliation{Department of Physics, Indian Institute of Technology Delhi, Hauz Khas, New Delhi:110016, India}
\email{saswata@physics.iitd.ac.in [SB], Arunima.Singh@physics.iitd.ac.in[AS]}
\phone{+91-2659 1359}
\fax{+91-2658 2037}

\title[An \textsf{achemso} demo]
{MoS{$_2$} and Janus (MoSSe) Based 2D van der Waals Heterostructures: Emerging Direct Z-scheme Photocatalysts$^\dag$}


\begin{document}






\begin{abstract} 
\noindent 
	Two-dimensional (2D) materials viz. transition metal dichalcogenides (TMD) and transition metal oxides (TMO) offer a platform that allows creation of heterostructures with a variety of properties. The optoelectronic industry has observed an upheaval in the research arena of MoS{$_2$} based van der Waals (vdW) heterostructures (HTSs) and Janus structures. Therefore, interest towards these structures are backed by the selectivity in terms of electronic and optical properties. The present study investigates the photocatalytic ability of bilayer, MoS{$_2$} and Janus (MoSSe) based vdW HTSs viz. MoS{$_2$}/TMO, MoS{$_2$}/TMD, MoSSe/TMO and MoSSe/TMD, by first-principles based approach under the framework of (hybrid) density functional theory (DFT) and many body perturbation theory (GW approximation). We have considered HfS{$_2$}, ZrS{$_2$}, TiS{$_2$}, WS{$_2$} and HfO{$_2$}, T-SnO{$_2$}, T-PtO{$_2$} from the family of TMD and TMO, respectively.  The photocatalytic properties of these vdW HTSs are thoroughly investigated and compared with the respective individual monolayers by visualizing their band edge alignment, electron-hole recombination and optical properties. Strikingly we observe that, despite most of the individual monolayers do not perform optimally as a photocatalyst, type II band edge alignment is noticed to vdW HTSs and they appear to be efficient for photocatalysis via Z-scheme. Moreover, these HTSs have also shown promising optical response in the visible region. Finally electron-hole recombination, H$_2$O adsorption and hydrogen Evolution Reaction (HER) results establish that MoSSe/HfS{$_2$}, MoSSe/TiS{$_2$}, MoS{$_2$}/T-SnO{$_2$}, MoS{$_2$}/ZrS{$_2$} and MoSSe/ZrS{$_2$} are probable, most efficient Z-scheme photocatalysts.
\end{abstract}
\section{Introduction}
\noindent In present context, solar energy is the supreme resource to combat energy and environment related issues. With an aim to enhance the utilization of solar energy, the field of photocatalysis for water splitting and pollutant degradation have gained interest~\cite{ni2007review,yao2015pollutant,opoku2017role,yu2019facile}. Further, the demand for exciting materials that serve selectivity, have directed the focus towards the 2D materials. Amongst these, MoS{$_2$} is one of the widely studied 2D materials that belongs to the transition metal dichalcogenides (TMD) family. The other classification constitutes graphene family, transition metal oxides (TMOs) and MXenes~\cite{gupta2015recent}. These materials have been vastly studied in the regimes of defect study~\cite{asjpcc, chen2018ACSNano}, photovoltaics~\cite{rao2015comparative,singh2019ACSApplInter} and optoelectronics~\cite{zeng2013optical,wang2015JPCC}. These widely studied (both theoretically and experimentally) TMDs are MoS{$_2$}, MoSe{$_2$}, WS{$_2$} and WSe{$_2$}. However, other TMDs such as HfS{$_2$}, ZrS{$_2$} and TiS{$_2$} have also been a matter of research for their salient electronic, vibrational and optical properties~\cite{glebko2018electronic,lau2019electronic,mattinen2019atomic}. Conversely, TMOs have not been established with similar level of experimental database as of TMDs. Leb\`{e}gue \textit{et al.}~\cite{lebegue2013two} indicate synthesis of TMOs as challenging due to the rare occurrence of their stable bulk phases. Only few of these materials have been synthesized experimentally~\cite{yang2019formation}. In this scenario, theoretical studies of the TMOs have showcased a huge potential, which helps in opening the scope of further research~\cite{leong2016two,liao2013new}. Another class of 2D materials that have recently drawn a lot of interest is Janus~\cite{ju2020janus,sun2018janus}. These materials, having the form of MXY (M = Mo and X,Y = O, S, Se and Te; X{$\neq$}Y), lack mirror symmetry and have vertical dipole~\cite{song2019suppressed,sun2020b2p6}. It exhibits superior photocatalytic activities amongst its counterparts~\cite{ma2018janus,luo2020mosse}. Engineering them for constructing van der Waals (vdW) heterostructures (HTSs) have established novel physics such as interlayer screening effects~\cite{kuc2015ChemSocRev, kumar2018interlayer,huang2017layer} and valley physics~\cite{terrones2013novel,zhang2018rashba,rivera2016valley}, etc. Therefore, the existing literatures indicate the aforementioned materials as promising in present research context~\cite{li2018multifunctional,li2019two,luo2019first,luo2019transition,ren2020high,wang2018mos}.
\begin{figure}[h]
	\centering
	\includegraphics[width=0.5\columnwidth,clip]{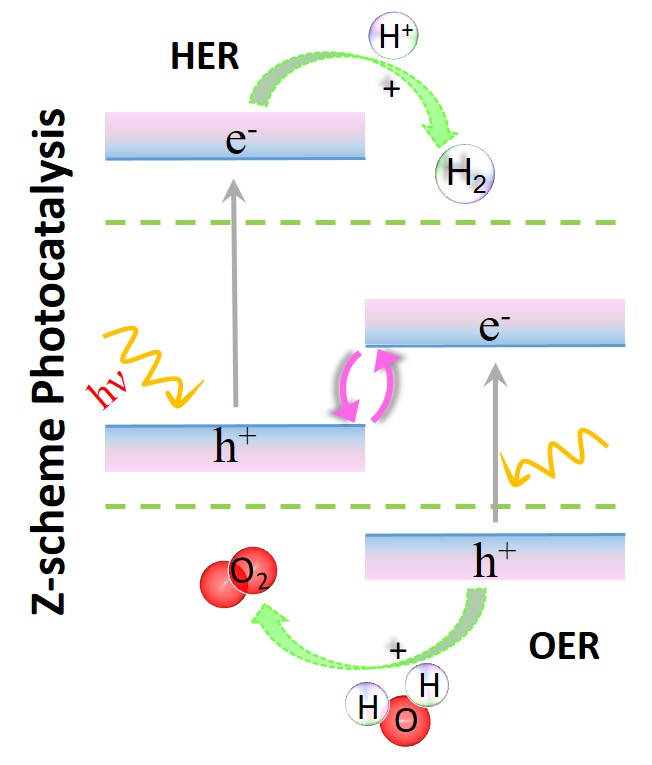}
	\caption{(Color online) Schematic of Z-scheme photocatalysis.}
	\label{fig:ZScheme}
\end{figure}
\\
\noindent Here we intend to select the vdW HTSs for the study of photocatalytic water splitting. Now, the MoS{$_2$} monolayer with direct band gap and apt redox potentials is a good photocatalyst~\cite{li2013single,parzinger2015photocatalytic}. However, its performance is affected by the electron (e{$^-$}) and hole (h{$^+$}) recombination as the carriers are generated in the same spatial region~\cite{tang2016spatial}. Therefore, an attempt in mitigating this issue is made by considering vdW HTSs, where due to the separation of e{$^-$} - h{$^+$} carriers in different layer (i.e. different spatial region), they are considered to be a successful alternative for photocatalytic water splitting~\cite{gao2019water,Direct_ZScheme_Zhang}. Here we have considered the class of vdW HTSs with type II band alignment for this application. Note that there are two types of photocatalysts in case of vdW HTSs (a) simple heterojunction photocatalysts and (b) Z-scheme systems. In former, Conduction Band Minimum (CBm) and Valence Band Maximum (VBM) of the vdW HTSs, straddle the redox potential~\cite{ren2019using,wang2018electronic}. Here CBm and VBM are on two different monolayers thereby facilitating spatial charge separation with reduction and oxidation reactions on different monolayer, respectively. But, to achieve sufficient redox ability for a specific reaction occurring on a single photocatalyst, a larger bandgap, that simultaneously inhibit e{$^-$} - h{$^+$} recombination and correlates to higher oxidation/reduction ability, is required. However, to enhance the efficiency, a smaller bandgap is desirable to increase the efficiency of light harvesting. Therefore, it is of a paramount importance for improving the photocatalytic performance, where Z-scheme~\cite{Direct_Z_Fu,liu2020situ,xia2019latest} brings the solution by manipulating the photogenerated e{$^-$} - h{$^+$} pairs as shown in the Z-scheme mode (see Fig.~\ref{fig:ZScheme}). Like before, here as well having type II alignment is mandatory but the CBm and VBM of the vdW HTSs do not straddle the redox potentials in the same manner as before. The layer with CBm straddling the reduction level and the another layer with VBM straddling the oxidation level are involved in hydrogen evolution reaction (HER) and oxygen evolution reaction (OER), respectively (see Fig.~\ref{fig:ZScheme}). The higher interlayer recombination of e{$^-$} - h{$^+$} as compared to that of intralayer is a necessary step in the process. Thus, photoabsorption is improved due to two different band gaps leading to HER and OER on different systems~\cite{maeda2013z, li2016z, bard1979photoelectrochemistry}. The weak vdW interlayer interaction in vdW HTSs facilitates lubricancy or interlayer movement that further assists the Z-scheme photocatalysis~\cite{Direct_ZScheme_Zhang}. Although, some vdW HTSs have been studied under this regime~\cite{lang2020phosphorus, Direct_ZScheme_Zhang, Direct_Z_Fu, ju2018dft}, no extensive study on MoS{$_2$} based vdW HTSs consisting of TMDs (viz. HfS{$_2$}, ZrS{$_2$}, TiS{$_2$}) and TMOs (viz. HfO{$_2$}, T-PtO{$_2$}, T-SnO{$_2$}) as second layer are observed so far. On a similar note, the database of Janus (MoSSe) based vdW HTSs for the photocatalytic application is at an initial stage~\cite{luo2020mosse,sun2018janus}. Hence, there is a justified interest to explore their applicability as photocatalysts. The MoSSe based vdW HTSs have further given rise to two possible stacking arrangements. Contrary to the structure of MoS{$_2$}, the MoSSe lacks mirror symmetry. Hence, it is crucial to compare the possible stacking configurations of MoSSe based vdW HTSs with the MoS{$_2$} based vdW HTSs, for the photocatalytic application.\\

\noindent In this article, we have presented an exhaustive comparative study of MoS{$_2$} and MoSSe based vdW HTSs, where the second layer is from the family of (i) TMDs viz. WS{$_2$}, HfS{$_2$}, ZrS{$_2$}, TiS{$_2$} and (ii) TMOs viz. HfO{$_2$}, T-PtO{$_2$}, T-SnO{$_2$}. The existing literatures so far, have not discussed the above listed vdW HTS configurations for the Z-scheme photocatalysis. Until date earlier reports exist only on monolayers of type II vdW HTSs of TMDs and TMOs for photocatalytic applications~\cite{rasmussen2015computational}. Here, using state-of-the-art theoretical methodologies within the framework of hybrid density functional theory (DFT) and manybody perturbation theory (MBPT) (viz. G$_0$W$_0$) we have systematically studied the important parameters for photocatalytic applications. First, we introduce the different possible stable configurations of vdW HTSs. Next we discuss the band edge alignment of 2D monolayers and their corresponding vdW HTSs. Subsequently, their recombination path is well analyzed to check capability as a Z-scheme photocatalyst. Finally, their accurate optical response is computed and understood with MBPT technique for consideration in photocatalytic devices. Strikingly we observe that, despite most of the individual monolayers do not perform optimally as a photocatalyst, type II band edge alignment is noticed to vdW HTSs and they appear to be efficient for photocatalysis via Z-scheme.

\section{Methodology}
\noindent We have employed first-principles based methodology under the framework of DFT~\cite{martin2004electronic, martin2016interacting, freysoldt2014RevModPhys, feng2014MaterChemPhys, hohenberg1964PhysRev, kohn1965PhysRev}. The PAW pseudopotentials are used in our calculations using plane wave basis set as employed in Vienna \textit{Ab initio} Simulation Package (VASP)~\cite{kresse1996efficient,blochl1994projector,blum2009ComputPhysCommun}. The exchange-correlation (xc) interaction amongst electrons are accounted by Generalized Gradient Approximation (GGA) with the functional form as proposed by Perdew-Burke-Ernzerhof (PBE)~\cite{stampfl1999PRB, perdew1996PRL}. The hybrid density functional has also been employed to account for the same with the functional proposed by Heyd-Scuseria-Ernzerhof (HSE06)~\cite{heyd2003JChemPhys}. The hybrid functional (HSE06) considers 25\% mixing {$(\alpha)$} of the short range Hartree-Fock (HF) exchange. Its long range part is described by GGA-PBE functional.\\
\noindent The vdW HTSs are optimized using PBE functional and HSE06 is used to determine single point energy. The conjugate gradient minimization is performed with the Brillouin Zone (BZ) sampling of {$2\times2\times1$} K-grid and the energetics are obtained by the BZ sampling of {$16\times16\times1$} K-grid. The energy tolerance of 0.001 meV and force tolerance of 0.001 eV/{\AA} have been used for optimization. In the ground state calculations, the plane wave cut-off energy is set to 600 eV. The vdW HTSs are modelled with 20 {\AA} vacuum in order to avoid electrostatic interactions between the periodic images. The two-body vdW interaction  as devised by Tkatchenko-Scheffler has been employed~\cite{tkatchenko2009PRL,tkatchenko2012PRL}. The correction parameter is based on Hirshfeld partitioning of the electron density. Note that, we have not included spin-orbit coupling (SOC) in our calculations, since the previous literatures have reported only slight change in band gap due to the same~\cite{Direct_Z_Fu,weng2018honeycomb,ren2020direct}. The optical properties are calculated using the GW approach~\cite{onida2002electronic,jiang2012electronic}. We have performed hybrid calculations
(HSE06), as an initial step for single shot GW calculations [i.e. G$_0$W$_0$@HSE06].

\section{Results and Discussions}
\subsection{Heterostructure stacking}
\noindent We have constructed commensurate bilayer vdW HTSs~\cite{rahman2018commensurate} with minimum lattice mismatch between the layers.  The vdW HTSs have been formed with monolayer of MoS{$_2$} (or MoSSe) along with that of WS{$_2$}, ZrS{$_2$}, HfS{$_2$}, TiS{$_2$}, HfO{$_2$}, T-PtO{$_2$} and T-SnO{$_2$}, in vertical alignment. The specifications of the monolayers (see lattice parameters in SI) corroborate with the existing literature~\cite{bastos2019ab,rasmussen2015computational,haastrup2018computational}. All the monolayers are constituted as {$2\times2\times1$} supercell (24 atoms) except ZrS{$_2$} and HfS{$_2$}, where {$\sqrt{3}\times\sqrt{3}\times1$} supercell consists of 21 atoms. Here, the stacking styles between two monolayers have not been varied in each vdW HTS. This is due to the fact that, the binding energy change of merely few meV is observed with the change in stacking styles of two monolayers in a particular vdW HTS~\cite{xia2018enhanced,hu2016stacking}.
\begin{figure}[h]
	\centering
	\includegraphics[width=1\columnwidth,clip]{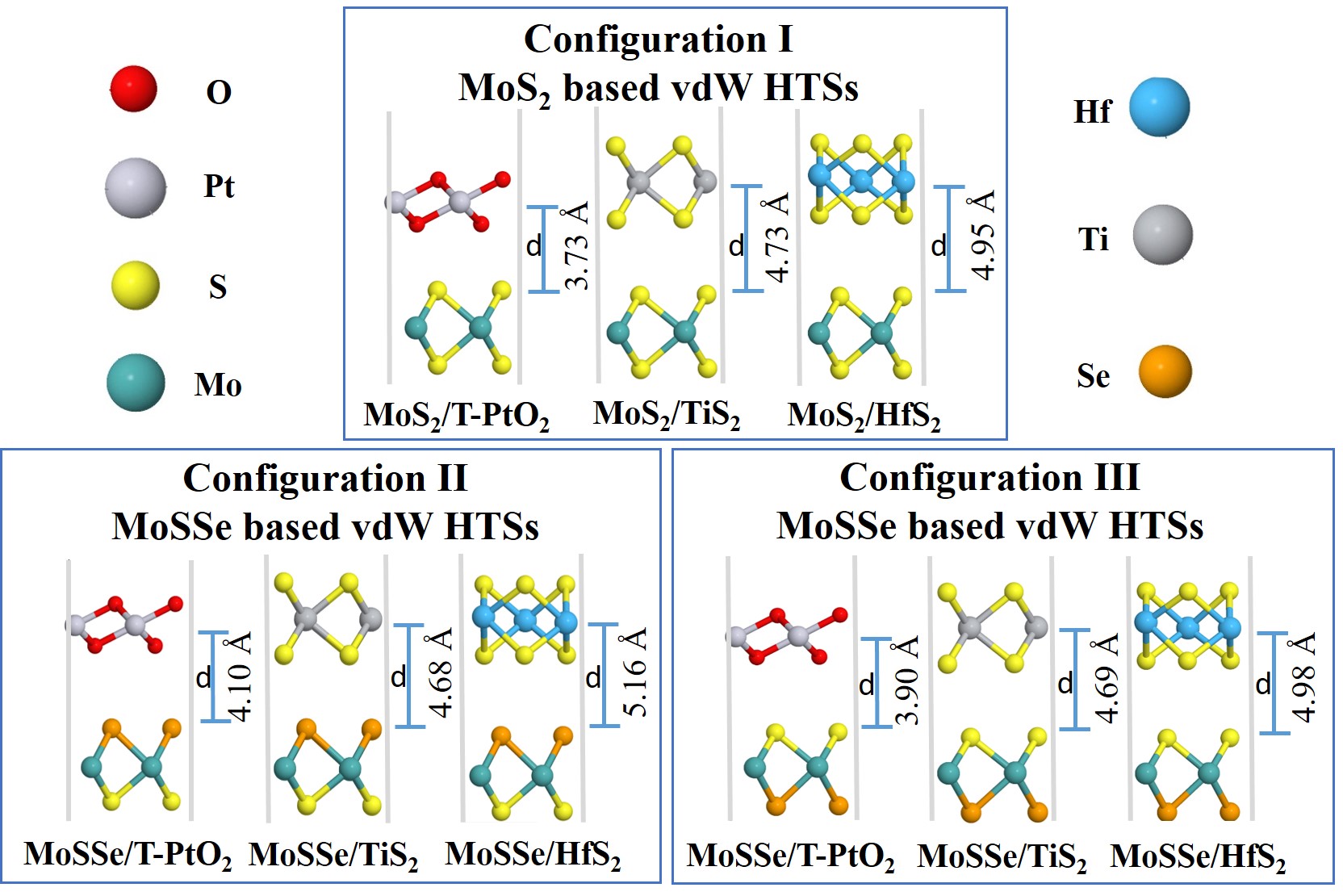}
	\caption{(Color online) vdW HTSs with configuration I, II and III. MoS{$_2$}/BX{$_2$} (configuration I), MoSSe/BX{$_2$} (configuration II), where Se atomic layer is at interface and MoSSe/BX{$_2$} (configuration III), where S atomic layer is at interface.}
	\label{fig:Stack}
\end{figure}

\noindent We have calculated the binding energy of vdW HTSs by the expression~\cite{xia2018enhanced},
\begin{equation}
	\textrm{E}{_\textrm{b}} = \textrm{E}(\textrm{vdW HTSs}) - \textrm{E}(\textrm{MoS}_2 \textrm{ or MoSSe}) - \textrm{E}(\textrm{BX}_2)
\end{equation}

\noindent where E{$_\textrm{b}$} is the binding energy of the vdW HTS  MoS{$_2$}/BX{$_2$} (or MoSSe/BX{$_2$}); E(MoS{$_2$}) is the energy of monolayer MoS{$_2$}; E(MoSSe) is the energy of monolayer MoSSe and E(BX{$_2$}) is the energy of monolayer BX{$_2$} (where B = W, Hf, Zr, Pt, Sn and X = S or O). \\
\begin{table}[h]
	\caption{\ The corresponding bilayer vdW HTS lattice mismatch and binding energies of different configurations (MoS{$_2$}/BX{$_2$}(I), MoSSe/BX{$_2$}(II and III)) are enlisted} 
	
	\begin{tabular}[c]{p{0.09\textwidth}cp{0.06\textwidth}cp{0.07\textwidth}cp{0.07\textwidth}cp{0.07\textwidth}}\hline
		\\[-1em]
		\multirow{2}{*}{BX{$_2$}}&\multirow{2}{*}{Lattice mismatch (\%) }&\multicolumn{3}{c}{Binding Energy (eV)}\\ 
		&  & I & II & III\\ \hline
		\\[-1em]
		WS{$_2$}  &   0.0   &   -1.04  & -1.05 & -1.02\\ 
		\\[-1em]
		ZrS{$_2$}   &   2.5   &   -0.49  & -0.54 & -0.52\\
		\\[-1em]
		HfS{$_2$}  &   3.9  &   -0.30  &-0.35 & -0.33\\ 
		\\[-1em]
		TiS{$_2$}  &  2.5   &   -0.45  &-1.00 & -0.94\\ 
		\\[-1em]
		HfO{$_2$}  &  1.2   &   -0.40 &-0.54 & -0.47\\ 
		\\[-1em]
		T-PtO{$_2$}  &   0.4   &   -0.83  &-0.63 & -0.68\\ 
		\\[-1em]
		T-SnO{$_2$}  &   1.8  &   -2.19  &-2.24 & -2.22\\ \hline
	\end{tabular}
	\label{Table1}
	
\end{table}
\noindent Fig.~\ref{fig:Stack} shows the three configuration of vdW HTSs as considered in the present work. Configuration I is of the type MoS{$_2$}/BX{$_2$}. The stacking type of MoS{$_2$}/T-SnO{$_2$}, MoS{$_2$}/ZrS{$_2$} and MoS{$_2$}/WS{$_2$}(or HfO{$_2$}) are same as that of MoS{$_2$}/T-PtO{$_2$}, MoS{$_2$}/HfS{$_2$} and MoS{$_2$}/TiS{$_2$}, respectively. On replacing MoS{$_2$} with MoSSe, we have constituted configuration II and III. The former consists of Se atomic layer of MoSSe at the interface, whereas the latter has S atomic layer at the interface. Configuration II and III also follow the same analogy as configuration I. Table~\ref{Table1} enlists the lattice mismatch and the binding energy of all the vdW HTS configurations. We define the mismatch as (l(MoS$_2$) - l(BX$_2$))/l(BX$_2$), where l is the lattice constant of MoS$_2$ and BX$_2$ respectively.

\subsection{Band edge alignment}

\begin{figure}[!ht]
	\centering
	(a){\label{fig:1a}\includegraphics[width=0.6\columnwidth,clip]{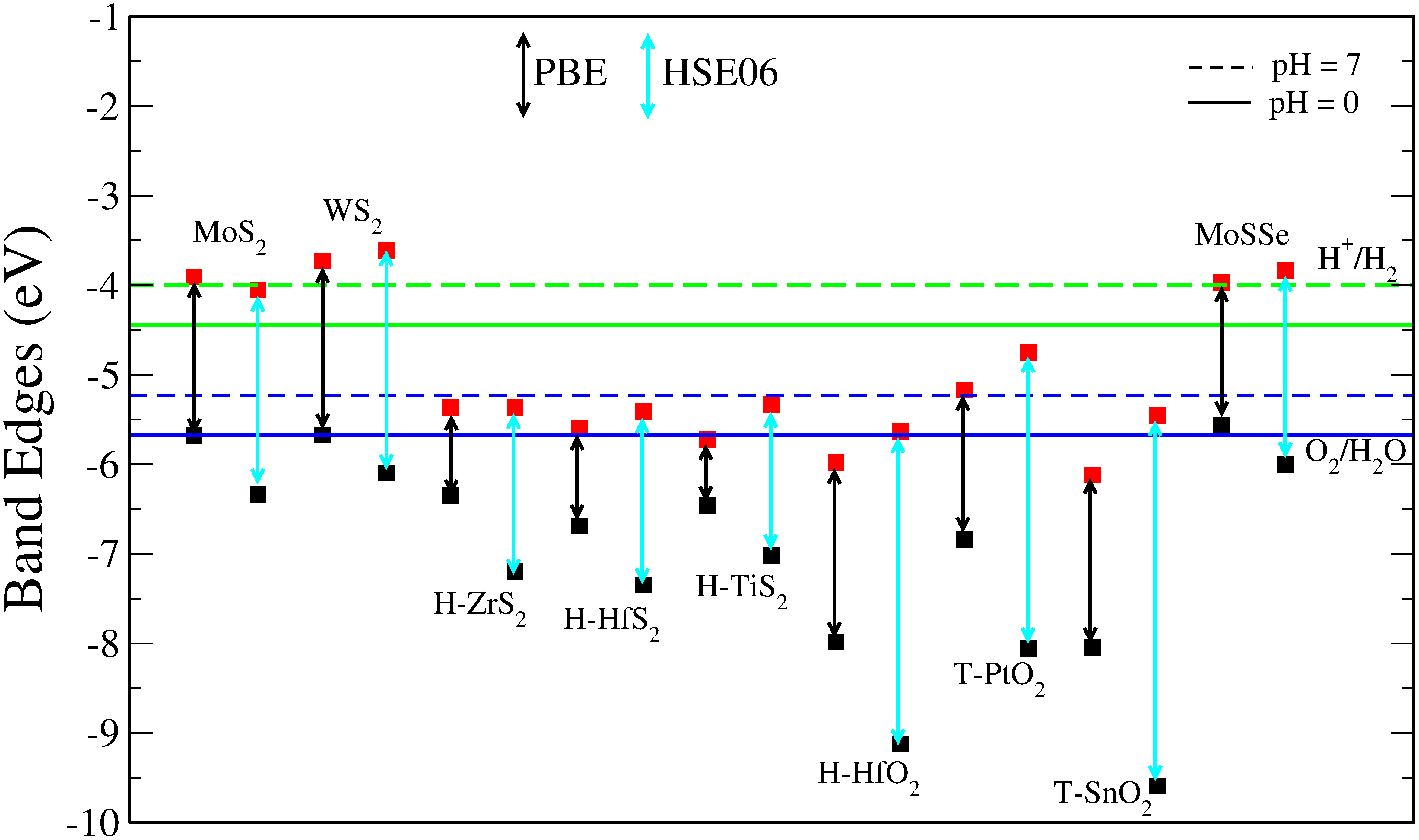}}
	(b){\label{fig:1b}\includegraphics[width=0.6\columnwidth,clip]{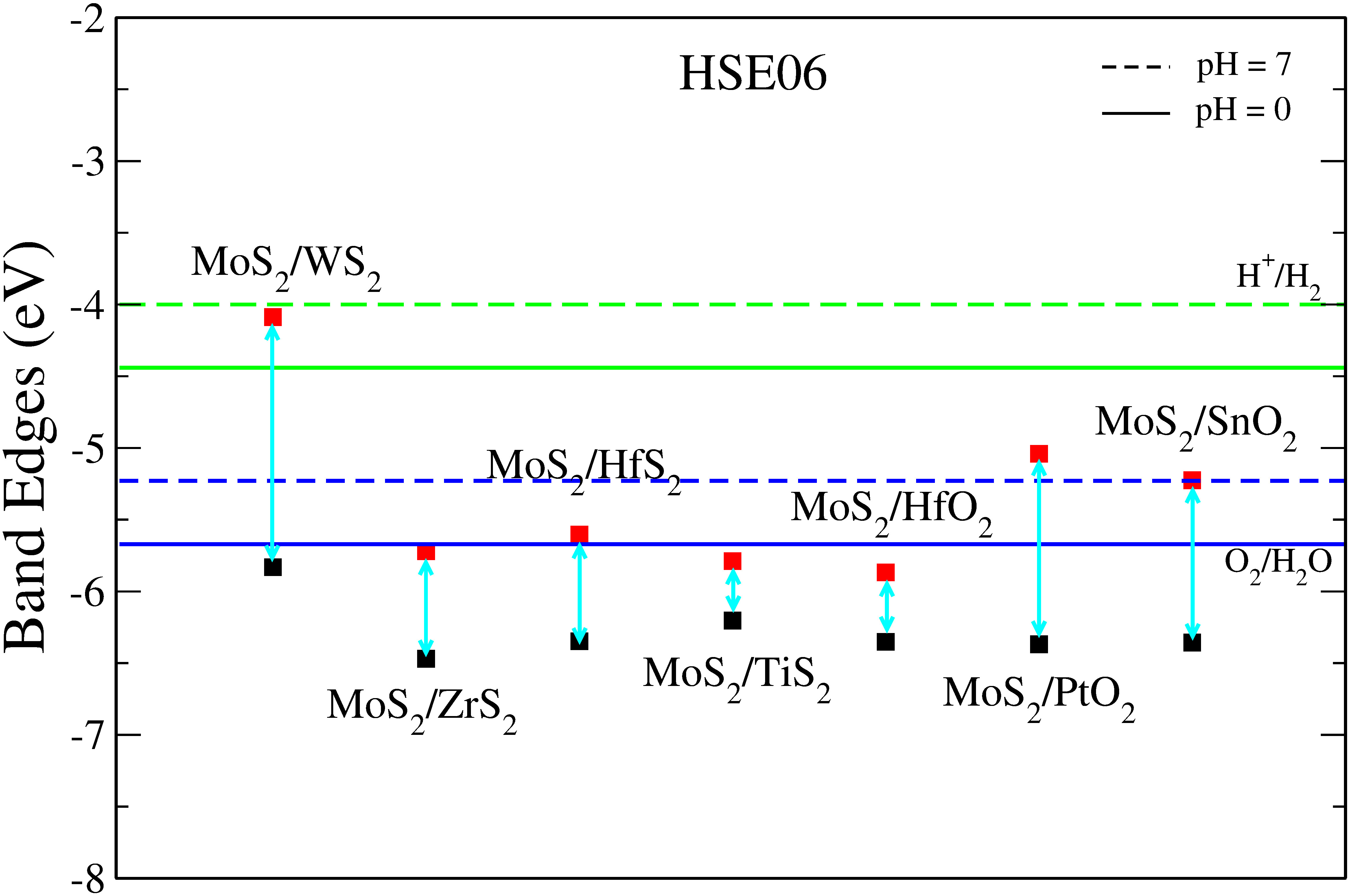}}
	(c){\label{fig:1c}\includegraphics[width=0.6\columnwidth,clip]{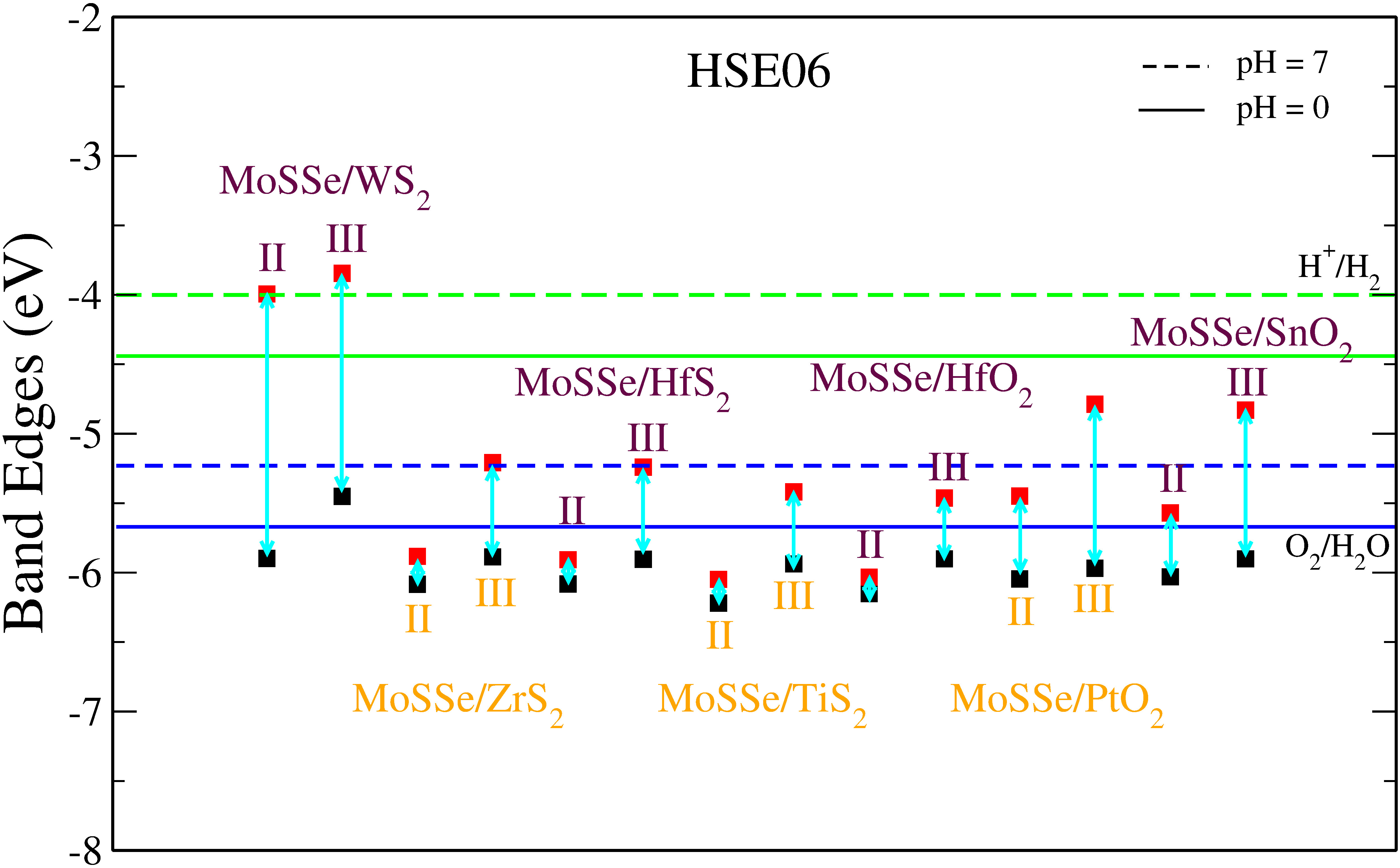}}
	\caption{(Color online) Band edge alignment with respect to water redox potentials of (a) individual monolayers, (b) MoS{$_2$}/BX{$_2$} vdW HTSs (configuration I), (c) MoSSe/BX{$_2$} vdW HTSs (configuration II and III), where BX{$_2$} refers to WS{$_2$}, ZrS{$_2$}, HfS{$_2$}, TiS{$_2$}, HfO{$_2$}, PtO{$_2$} and SnO{$_2$}. }
	\label{fig:1}
\end{figure}

\noindent The photocatalytic applications require the understanding of band gaps and absolute band edge positions. We have calculated these with both PBE and HSE06 functionals. Fig.~\ref{fig:1}(a) provides band edge alignment of the monolayers. We observe here that the chosen monolayers for vdW HTSs have type II heterojunction. In the Fig.~\ref{fig:1}, H{$^+$}/H{$_2$} and O{$_2$}/H{$_2$}O correspond  to the reduction and oxidation potential of water splitting, respectively, both for pH(0) (solid line) and pH(7) (dashed line). Fig.~\ref{fig:1}(a) shows that HSE06 and PBE have sufficient discrepancies in estimating band edge positions. Calculations with HSE06 functional incorporate HF exact exchange term resulting due to self-interaction error of e{$^-$}, which is not well taken care of in the case of PBE functional. Hence, the results presented in further plots are carried out by the HSE06 functional. 

\noindent Fig.~\ref{fig:1}(b) and Fig.~\ref{fig:1}(c) depict the absolute band edge positions of the MoS{$_2$} and MoSSe based vdW HTSs, respectively. Fig.~\ref{fig:1}(b) indicates that individual band alignment approximately predicts the combined alignment of the vdW HTSs. The weak vdW interaction between 2D monolayers is attributed to the aforementioned indication. Here, the type II alignment can be seen with VBM of MoS{$_2$} (or MoSSe ) monolayer and CBm of HfS{$_2$}, ZrS{$_2$}, TiS{$_2$}, HfO{$_2$}, T-PtO{$_2$} and T-SnO{$_2$} monolayers. The band gaps of the monolayers and vdW HTSs are listed in SI (Table I). Now, the CBm and VBM should lie a few eVs above and below the redox potentials for the material to be good for photocatalysis. However, in our case the vdW HTS band gap does not straddle the photocatalytic water splitting potentials. Hence, we analyze these bilayer vdW HTSs for their applicability as Z-scheme photocatalyst. The vdW HTSs deduced for further analysis from Fig.~\ref{fig:1} are the configurations consisting HfS{$_2$}, ZrS{$_2$}, TiS{$_2$}, HfO{$_2$}, T-PtO{$_2$} and T-SnO{$_2$}. Note that MoS{$_2$} (or MoSSe) monolayer (CBm) does form type II alignment with WS{$_2$} (VBM) as well and the HTS is observed to be capable for photocatalysis (but not via Z-scheme). However, we have included this in our manuscript as a reference, with an intention to showcase clear distinction w.r.t Z-scheme photocatalysts. Moreover, on observing Fig.~\ref{fig:1}(c) a unanimously similar trend infers that the band gap of vdW HTSs of configuration II is significantly different than that of configuration III. 

\begin{figure}[h]
	\centering
	\includegraphics[width=0.77\columnwidth,clip]{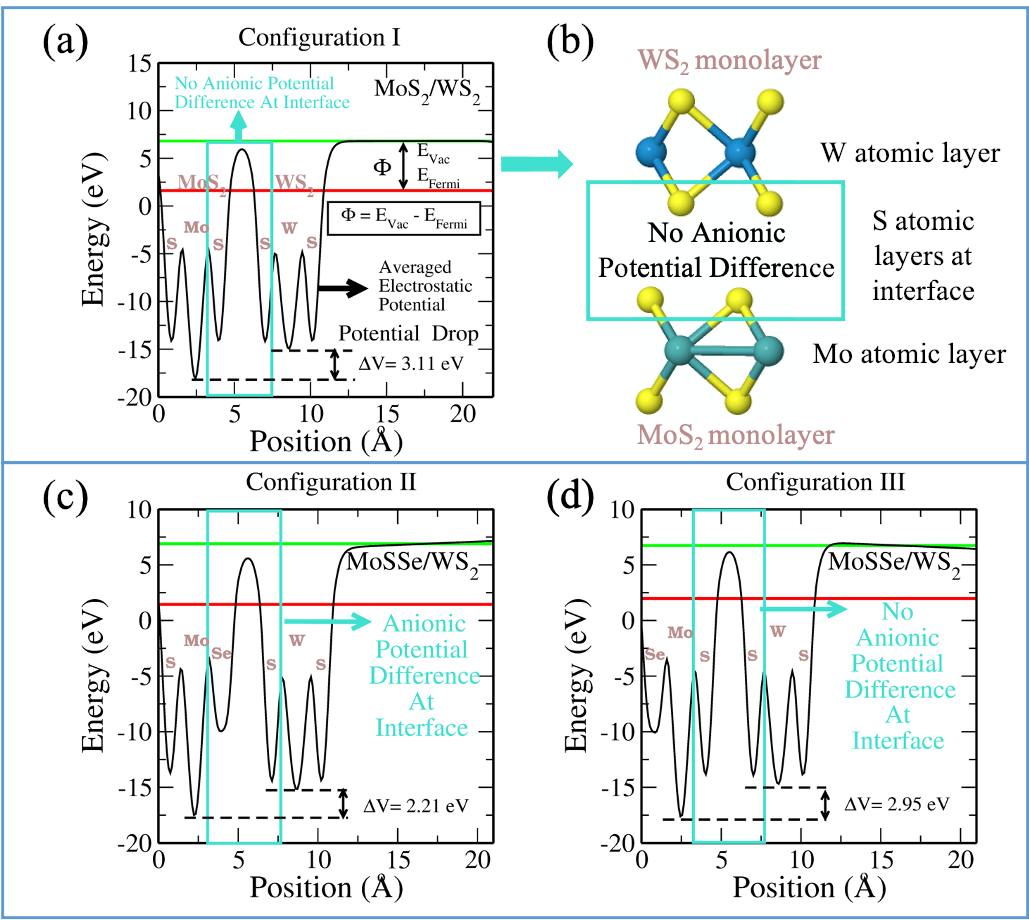}
	\caption{(Color online) Electrostatic potential plot of MoS{$_2$}/WS{$_2$} depicting (a) the  potential difference, (b) the work function in configuration I, (c) cationic and anionic potential difference in configuration II, and (d) configuration III.}
	\label{fig:exp_wf}
\end{figure}

\noindent It should be noted that the absolute band edge alignment is obtained with respect to the zero vacuum level. The associated E{$_\textrm{Vac}$} (vacuum level) of each calculation is obtained from the electrostatic potential of the system. As is evident from the term, the plot explains the electrostatic potential corresponding to each atomic layer (see Fig.~\ref{fig:exp_wf}). It is pertinent to understand the plot as it informs about the work function and potential difference between the monolayers. Thus, we are able to analyze the interlayer charge transfer that indicates the photocatalytic strength. Fig.~\ref{fig:exp_wf} corresponds to the electrostatic potential plot of MoS{$_2$}/WS{$_2$}. Considering left to right chronology in Fig.~\ref{fig:exp_wf}(a), we observe the potential of S, Mo, S atomic layers of  MoS{$_2$} monolayer and S, W, S atomic layers of WS{$_2$} monolayer. We see a clear difference in the potentials of cationic layers viz. Mo and W atomic layers. The gradient between the monolayers facilitate the charge separation between them. Fig.~\ref{fig:exp_wf}(b) shows the corresponding MoS{$_2$}/WS{$_2$} vdW HTS describing the respective atomic layers. \\
\\
Fig.~\ref{fig:exp_wf}(c) and (d), show the plots for MoSSe/WS{$_2$} configuration II and III, respectively. We observe, that in configuration II (Fig.~\ref{fig:exp_wf}(c)) there is an additional potential gradient between Se atomic layer of MoSSe and S atomic layer of the WS{$_2$} at the interface, along with the cationic potential difference. In case of configuration III, the interface has no anionic gradient (due to presence of S in both the layers facing each other) and hence, charge separation is mainly the result of cationic gradient. Therefore, we see different band gaps in configuration III and configuration II. Further, one should always remember, that these configurations are type II and the band edge levels are on two different layers. Therefore, any potential gradient at interface will definitely effect the corresponding band edge levels. Fig. S1-S6 in SI discuss the electrostatic potential for other vdW HTSs. Fig. S4 explains the change in band gap on the basis of dipole direction. Dipole direction is seen from S to Se i.e. from lower to higher potential. So, we observe opposite dipole directions in configurations II and III, thereby affecting the interfacial interactions. Thus, the vdW HTSs with TMOs are also inline with the above discussion and inference.\\
\begin{table}[h]
	\caption{\ The Hirshfeld charge and work function of different configurations (I: MoS{$_2$}/BX{$_2$}, II: MoSSe/BX{$_2$} (Se interfacial layer), III: MoSSe/BX{$_2$} (S interfacial layer))} 
	
	\begin{tabular}[c]{p{0.09\textwidth}p{0.06\textwidth}p{0.06\textwidth}p{0.06\textwidth}p{0.06\textwidth}p{0.06\textwidth}p{0.06\textwidth}}\hline
		\\[-1em]
		\multirow{2}{*}{BX{$_2$}}&\multicolumn{3}{c}{Hirshfeld charge (e) }&\multicolumn{3}{c}{Work function ($\phi$) \textrm{(eV)}}\\
		&I&II&III&I&II&III\\
		\hline
		\\[-1em]
		WS{$_2$}  &   0.64   &   0.19  & 3.00 & 5.19&5.34&4.76\\ 
		\\[-1em]
		ZrS{$_2$}   &   -0.26   &   -0.25  & -0.16 & 5.98&6.02&5.56\\ 
		\\[-1em]
		HfS{$_2$}  &   -0.08  &   -0.14  &-0.07 & 5.98&6.01&5.58\\ 
		\\[-1em]
		TiS{$_2$}  &  -0.09   &   -0.25  &-0.12 & 6.02&6.15&5.69\\ 
		\\[-1em]
		HfO{$_2$}  &  -0.04   &   -0.12 &-0.09 & 6.12&6.13&5.70\\ 
		\\[-1em]
		T-PtO{$_2$}  &   0.03   &   -0.08  &-0.04 & 5.78&5.75&5.53\\
		\\[-1em]
		T-SnO{$_2$}  &   -0.10  &   -0.16  &-0.06 & 5.91&5.79&5.46\\ \hline
	\end{tabular}
	\label{Table2}	
\end{table}

\noindent We have also calculated the associated charge density on the layers and the work function of the vdW HTSs. The work function ($\phi$) is significant parameter to understand the charge separation or transfer at the interface. It is defined as follows:
\begin{equation}
	\phi = \textrm{E}_\textrm{Vac} - \textrm{E}_\textrm{Fermi}, 
\end{equation}
\noindent where E{$_\textrm{Vac}$} and E{$_\textrm{Fermi}$} are the electrostatic potential corresponding to the vacuum level and Fermi level, respectively (see Fig.~\ref{fig:exp_wf}). Table~\ref{Table2} gives the total Hirshfeld charge (of all the atoms in the monolayer of the HTS) and the work function of vdW HTSs. The estimated values of the work function of free standing monolayer ($\phi_\textrm{M}$ as in Table~\ref{Table3}), have been corroborated with the existing literature~\cite{haastrup2018computational}\footnote{Note that the mentioned reference has considered 15 {\AA} vacuum, which we have found to be converged at 20 {\AA} vacuum. The slight difference (i.e. not exact match) may be due to this change.}. The Hirshfeld charge in case of configuration II show greater charge transfer than that of as in the case of configuration I and III. This is attributed to the anionic and cationic potential difference at the interface. The lesser charge transfer corresponds to the lower work function in configurations I and III. The plane averaged charge density difference $\Delta \rho$ has been calculated (see section V in SI) by:
\begin{equation}
	\Delta \rho = \rho(\textrm{vdW HTSs}) - \rho(\textrm{MoS}_2 \textrm{ or MoSSe}) - \rho(\textrm{BX}_2)
\end{equation}
\noindent where {$ \rho(\textrm{vdW HTSs})$}, {$ \rho(\textrm{MoS}_2) $}, {$ \rho(\textrm{MoSSe}) $} and {$ \rho(\textrm{BX}_2) $}  are the charge density of the vdW HTS, monolayer MoS{$_2$}, monolayer MoSSe and monolayer BX{$_2$}, respectively.
\begin{table}[h]
	\caption{\ The estimated band bending between vdW HTSs and free-standing monolayer ($\phi_\textrm{M}$ is the work function of free standing monolayer, I: MoS{$_2$}/BX{$_2$}, II: MoSSe/BX{$_2$} (Se interfacial layer), III: MoSSe/BX{$_2$} (S interfacial layer))} 
	\begin{center}
		
		\begin{tabular}[c]{p{0.09\textwidth}p{0.10\textwidth}p{0.07\textwidth}p{0.07\textwidth}p{0.07\textwidth}}\hline
			\\[-1em]
			\multirow{2}{*}{BX{$_2$}}&\multirow{2}{*}{$\phi_\textrm{M}$ (eV)}&\multicolumn{3}{c}{{$\Delta \textrm{E}_\textrm{BB} $} \textrm{(eV)}}\\
			&&I&II&III\\
			\hline
			\\[-1em]
			WS{$_2$}  &   5.53   &   -0.34&-0.19&-0.77\\ 
			\\[-1em]
			ZrS{$_2$}   &   6.95   & -0.97&-0.93&-1.39\\ 
			\\[-1em]
			HfS{$_2$}  &   7.13  &   -1.15&-1.12&-1.55\\ 
			\\[-1em]
			TiS{$_2$}  &  7.01  &   -0.99&-0.86&-1.32\\ 
			\\[-1em]
			HfO{$_2$}  &  8.86  &   -2.74&-2.73&-3.16\\ 
			\\[-1em]
			T-PtO{$_2$}  &   7.86   & -2.08&-2.11&-2.33\\ 
			\\[-1em]
			T-SnO{$_2$}  &   9.06  &  -3.15&-3.27&-3.60\\ \hline
		\end{tabular}
		\label{Table2b}
	\end{center}
\end{table}

\noindent We can understand the charge transfer by calculating the band bending ($\Delta \textrm{E}_\textrm{BB}$) as well. This parameter is estimated by the Fermi-level difference between the vdW HTSs and the corresponding free-standing monolayer, with the expression as follows~\cite{padilha2015van}:
\begin{equation}
	\Delta \textrm{E}_\textrm{BB} = \phi - \phi_\textrm{M} 
\end{equation}
\noindent where $\phi$ is the work function of vdW HTSs and $\phi_\textrm{M}$ is the work function of free standing monolayer. The $\Delta \textrm{E}_\textrm{BB}  < 0$ indicates charge transfer from the vdW HTSs to the monolayer. All the vdW HTSs, display $\Delta \textrm{E}_\textrm{BB} < 0$ i.e. charge gained by the monolayer (Table~\ref{Table2b}). The observation is in sync with the calculated Hirshfeld charges as noted in Table~\ref{Table2}. It is to be noted that in case of WS{$_2$}, $\Delta \textrm{E}_\textrm{BB} $ is more negative with $\phi_\textrm{M}$(MoS$_2$) = 5.82 eV than that with $\phi_\textrm{M}$(WS$_2$). Hence, the associated Hirshfeld charge on the WS$_2$ monolayer is positive. The above observations indicate MoSSe based configurations (II and III) to play a crucial role in photocatalytic applications.\\
\noindent The aforementioned discussions along with Fig.~\ref{fig:1}(a) have showcased the spatial charge separation due to the work function difference. Therefore, with the e{$^-$} - h{$^+$} pair generation in the individual monolayers of vdW HTSs, the MoS$_2$ (or MoSSe) monolayer will be positively charged as it contributes to the VBM of vdW HTS, which will restrict the flow of e{$^-$} to other layer (BX$_2$). Moreover, BX$_2$ monolayer that contributes to the CBm of vdW HTS, does not straddle the reduction potential. Hence, MoS$_2$ (or MoSSe) monolayer will observe restricted e{$^-$} motion from its CBm to BX$_2$ CBm, thereby facilitating the e{$^-$} - h{$^+$} recombination path of BX$_2$ CBm to MoS$_2$ VBM and promoting the standard Z-scheme mechanism.

\subsection{Recombination in vdW HTSs}
\noindent Considering the case of photocatalysts that straddle the potentials, lesser recombination is synchronous with the effective photocatalytic capability. However, vdW HTSs for applicability as Z-scheme potocatalysts, require higher recombination in them as compared to their constituent monolayers. Hence, the layer with CBm above the reduction potential can facilitate HER, while, the layer with VBM below the oxidation potential can facilitate OER. In view of this, we have analyzed the recombination rate, which is indicated by the effective mass of electrons and holes. The effective mass has been calculated by the parabolic fitting at CBm and VBM in the bandstructure obtained by the hybrid calculations~\cite{ganose2018sumo}. The bandstructure of MoS{$_2$}/ZrS{$_2$} has been found to be previously reported~\cite{lu2017impact} and is in sync with that reported in the present work. The similar database of other vdW HTSs discussed here are previously unreported. 
\begin{table}[h]
	\caption{\ The effective mass ratio (D) of the monolayers (BX{$_2$}) except  MoS{$_2$} and MoSSe (both have same value of D = 1.21)} 
	\begin{center}
		\begin{tabular}[c]{p{0.04\textwidth}p{0.06\textwidth}p{0.06\textwidth}p{0.06\textwidth}p{0.06\textwidth}p{0.06\textwidth}p{0.09\textwidth}p{0.09\textwidth}}\hline 
			\\[-1em]
			BX{$_2$}&WS{$_2$}&ZrS{$_2$}&HfS{$_2$}&TiS{$_2$}&HfO{$_2$}&T-PtO{$_2$}&T-SnO{$_2$} \\ \hline
			\\[-1em]
			D&1.29&1.58&1.17&0.03&5.88&4.66&9.00 \\  \hline
		\end{tabular}
		\label{Table3a}
	\end{center}
\end{table}
\begin{table}[h]
	\caption{\ The effective mass ratio (D) of the  vdW HTSs (I: MoS{$_2$}/BX{$_2$}, II: MoSSe/BX{$_2$} (Se interfacial layer), III: MoSSe/BX{$_2$} (S interfacial layer))} 
	\begin{center}
		\begin{tabular}[c]{p{0.15\textwidth}p{0.07\textwidth}p{0.07\textwidth}p{0.07\textwidth}}\hline 
			\\[-1em]
			\multirow{2}{*}{BX{$_2$}}&\multicolumn{3}{c}{Effective Mass Ratio (D)}\\
			&I&II&III \\  \hline
			\\[-1em]
			WS{$_2$}  &   2.50  & 0.85 &10.19\\
			\\[-1em]
			ZrS{$_2$} &   0.13  & 0.64 &0.62\\ 
			\\[-1em]
			HfS{$_2$} &   0.65  & 0.87 &0.71\\ 
			\\[-1em]
			TiS{$_2$} &   0.50  & 0.20&0.80\\ 
			\\[-1em]
			HfO{$_2$} &   0.32  & 0.34 &0.34\\ 
			\\[-1em]
			T-PtO{$_2$} &   0.49  & 0.50 & 0.49\\ 
			\\[-1em]
			T-SnO{$_2$} &   0.91  & 1.15 & 1.14 \\  \hline
		\end{tabular}
		\label{Table3}
	\end{center}
\end{table}

\noindent The recombination in the system can be estimated by relative ratio (D) that is defined as follows~\cite{zhang2012D}:
\begin{equation}
	\textrm{D} = \textrm{m}_\textrm{h}^{*} / \textrm{m}_\textrm{e}^{*} 
\end{equation}
The higher the variance of D from 1, the lesser is the recombination. Since, D refers to the relative ratio of the effective masses of hole ($\textrm{m}_\textrm{h}^{*}$) and electron ($\textrm{m}_\textrm{e}^{*}$), its value closer to 1 (i.e low variance) is ideal for high recombination in the system. This is obvious as value closer to 1 indicates similar electron and hole effective masses, resulting in easier recombination. 
On the contrary, large difference in the effective masses pose significant difference in carrier mobilities leading to the separation of carriers and hindrance in their recombination. Consider, Table~\ref{Table3} that enlists the value of D for monolayers and their corresponding vdW HTSs. The effective mass of carriers at CBm (i.e. $\textrm{m}_\textrm{e}^{*}$) and VBM (i.e. $\textrm{m}_\textrm{h}^{*}$) have been obtained along the designated direction in the high symmetry path $\Gamma$ - M - K - $\Gamma$ (see bandstructures i.e. Fig. S7 in SI). 
Also, it should be noted that the parameter is calculated from the bandstructures of the supercell than the primitive cells. Since, the supercell is small, the density of the bands near the VBM and CBm is not very high and hence, the band folding is not an issue here.\\
\noindent  Further, since we are considering Z-scheme photocatalysis, we require monolayer with lesser recombination as compared to that of their corresponding vdW HTSs. In order to review this we have checked the variance of D values from 1 for monolayers and vdW HTSs. Therefore, the vdW HTSs with lesser variance as compared to that of their respective monolayer can be considered probable for Z-scheme photocatalysis. As per the discussed approach, we have deduced MoSSe/ZrS$_2$ (configuration II and III), MoSSe/HfS$_2$ (configuration II),  MoSSe/TiS$_2$ (configuration III), MoS$_2$/SnO$_2$ and MoSSe/SnO$_2$ (configuration II and III) as the probable Z-scheme vdW HTSs. Amongst these, the MoS$_2$/SnO$_2$, MoSSe/HfS$_2$ (configuration II) and MoSSe/TiS$_2$ (configuration III) have shown least variance from 1. Their respective carrier mobilities~\cite{dai2015titanium} are computed as 46$\times$10$^2$ cm$^2$V$^{-1}$s$^{-1}$, 21$\times$10$^2$ cm$^2$V$^{-1}$s$^{-1}$ and 2$\times$10$^2$ cm$^2$V$^{-1}$s$^{-1}$. Additional details are provided in the supplementary material. Note that, we have also estimated their exciton binding energy (E{$_\textrm{B}$}), which showed comparatively smaller E{$_\textrm{B}$} of vdW HTSs as compared to that of MoS$_2$ and MoSSe monolayers (see section VII in SI). It should be mentioned here that the value D is an indicative approach and should not be considered as the sole criteria for e{$^-$} - h{$^+$} recombination. This yields future scope to explore more on carrier dynamics to understand the excitonic physics in these materials. We have, in addition, nevertheless, complied to the inferences with the study of optical response, carrier mobility, H$_2$O adsorption and HER.

\subsection{Absorption spectra}
\begin{figure}[!h]
	\centering
	\includegraphics[width=0.80\columnwidth,clip]{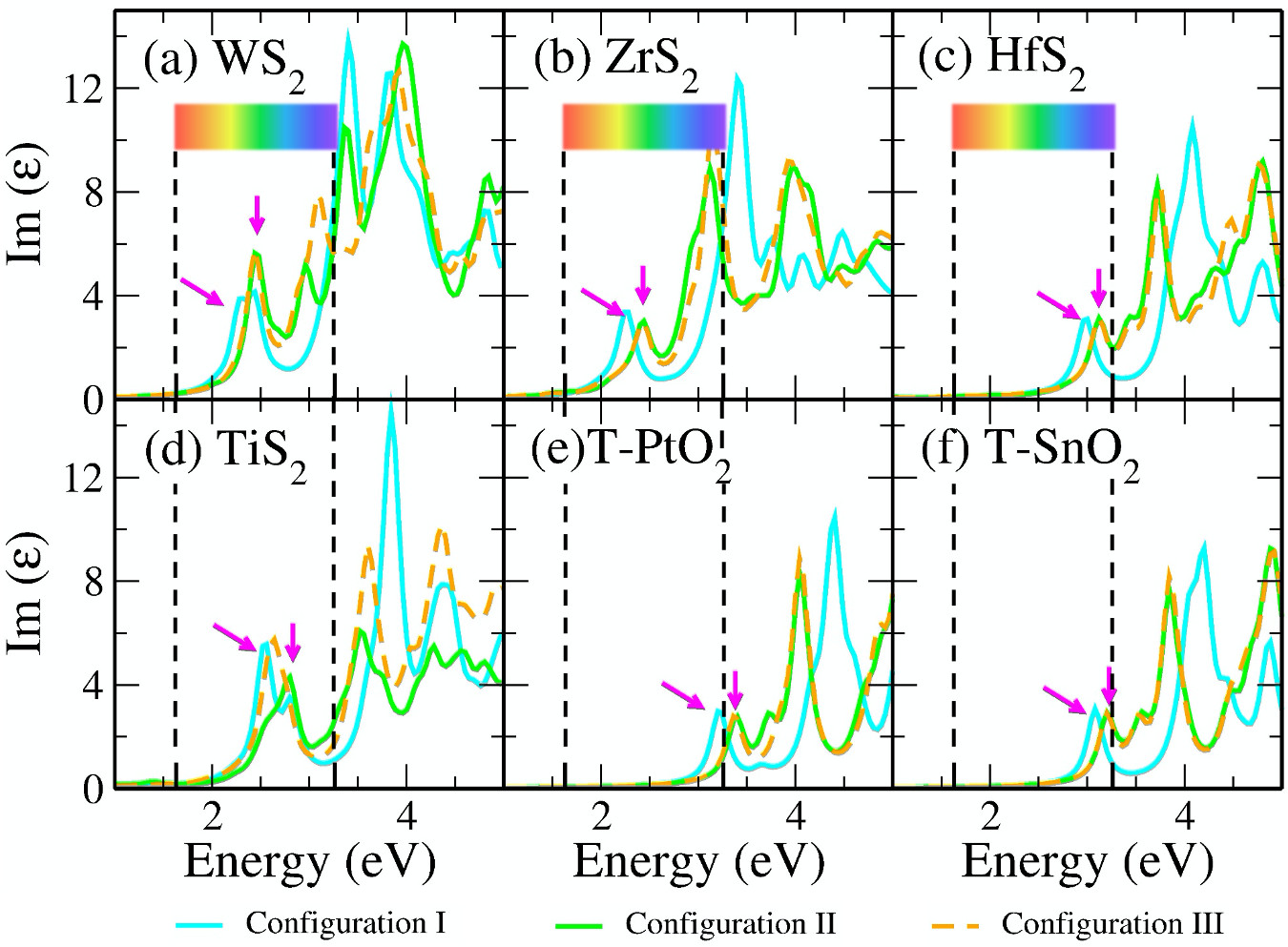}
	\caption{(Color online) Imaginary part of dielectric function vs. energy plot depicting the optical response of the vdW HTS configurations I, II and III by G$_0$W$_0$@HSE06. Here, the arrow indicates the initial peak position that is associated with the absorption onset.  }
	\label{fig:3}
	
\end{figure}
\noindent Optical response is amongst the important parameters involved in constructing the photocatalytic devices. Therefore, we have obtained the same for previously discussed vdW HTSs, that can be considered for the photocatalytic applications. Fig.~\ref{fig:3} gives the absorption spectra obtained from imaginary part of the dielectric function as computed by GW method (G$_0$W$_0$@HSE06). The dielectric function is a frequency dependent complex function ($\epsilon(\omega)$ = Re($\epsilon$) + Im($\epsilon$)) where the real part (Re($\epsilon$)) is computed from the Kramers-Kronig relation and the imaginary part (Im($\epsilon$)) is obtained from the interband matrix elements in the momentum space~\cite{basera2019self}. Since, the interband process consists of transitions to the unoccupied orbitals, the Im($\epsilon$) gives the absorption spectra. The first peak will, therefore, indicate the onset of absorption process. We observe that the vdW HTSs of MoS{$_2$}/TMD and MoS{$_2$}/TMO have response in the visible region. The spectra corresponding to MoSSe vdW HTSs (configuration II and III) have shown minute blue shift as compared to that of MoS{$_2$} vdW HTSs. Further, we observe from Fig.~\ref{fig:3}(b) that the vdW HTSs with ZrS{$_2$} and TiS{$_2$} monolayer show red shift as compared to other vdW HTSs (i.e., Fig.~\ref{fig:3}(a), (c), (d)-(f)). The spectra, thus, indicate the use of MoS{$_2$}/ZrS{$_2$}, MoSSe/ZrS{$_2$}, MoS{$_2$}/TiS{$_2$} and MoSSe/TiS{$_2$} for harnessing the overlap region of Im($\epsilon$) with the solar spectrum in order to construct the photocatalytic devices.

\subsection{H{$_2$}O Adsorption }
\begin{figure}[h]
	\centering
	\includegraphics[width=0.6\columnwidth,clip]{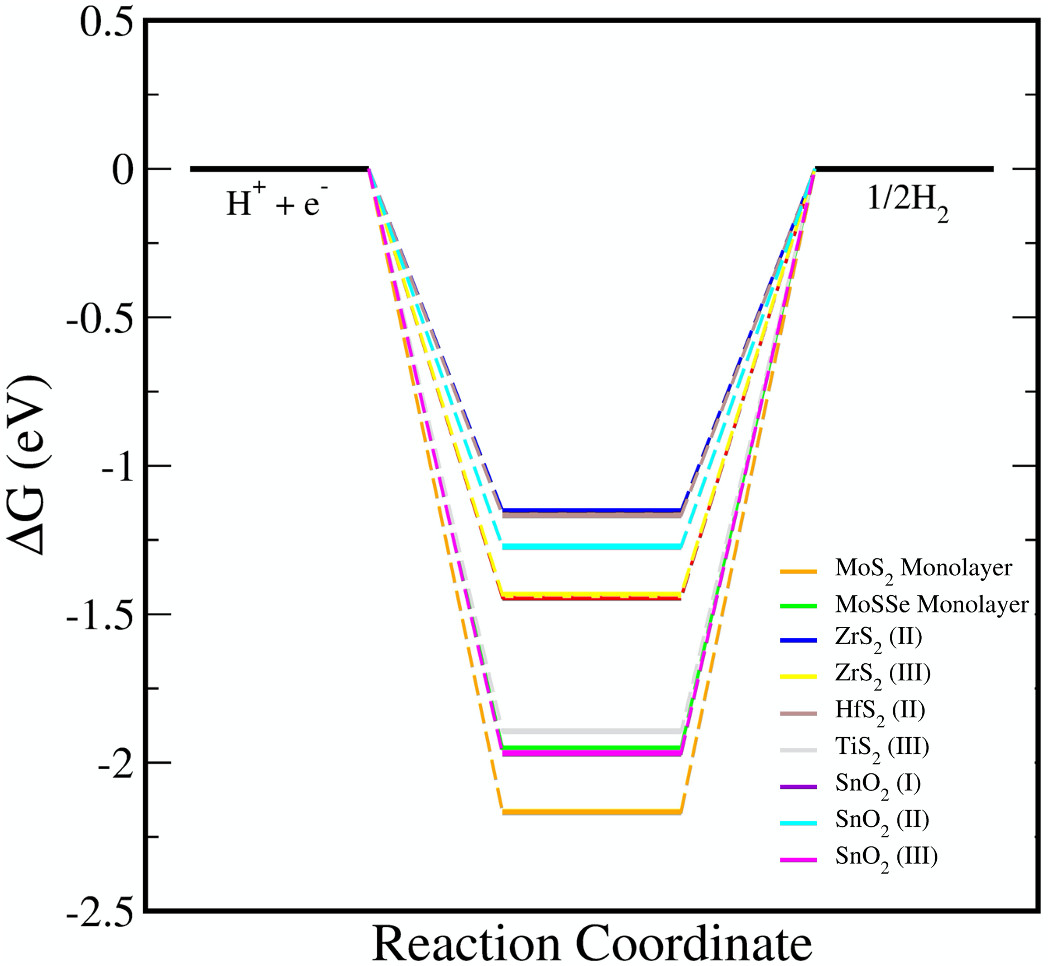}
	\caption{(Color online) The HER reaction free energy diagrams on monolayer and vdW HTSs at an electrode potential E{$_\textrm{SHE}$} = 0 V and pH = 0. At zero potential, (H{$^+$} + e{$^-$} ) can be expressed as 1/2 H{$_2$}.  }
	\label{fig:4}
\end{figure}

\begin{table}[h]
	\caption{\ H{$_2$}O adsorption on the monolayer surface} 
	\begin{center}
		\begin{tabular}[c]{p{0.09\textwidth}p{0.15\textwidth}p{0.15\textwidth}}\hline 
			\\[-1em]
			BX{$_2$}  & E{$_\textrm{ads}$} (eV)& Distance (\AA) \\ \hline
			\\[-1em]
			WS{$_2$}  &-0.09&2.88   \\ 
			\\[-1em]
			ZrS{$_2$} &-0.24&2.59  \\ 
			\\[-1em]
			HfS{$_2$} &-0.23&2.75   \\ 
			\\[-1em]
			TiS{$_2$} &-0.25&2.74   	  \\ 
			\\[-1em]
			T-PtO{$_2$} &-0.28&2.08   \\ 
			\\[-1em]
			T-SnO{$_2$} &-0.29&2.31    \\ \hline
			
		\end{tabular}
		\label{Table4}
	\end{center}
\end{table}
\noindent We have further calculated the H{$_2$}O adsorption in these vdW HTSs to confirm that the adsorption is supported for further redox reactions. The associated expression is as given:
\begin{equation}
	\textrm{E}{_\textrm{ads}} = \textrm{E}(\textrm{adsorbed H}_\textrm{2}\textrm{O}) - \textrm{E}(\textrm{vdW HTSs}) - \textrm{E}(\textrm{H}_\textrm{2}\textrm{O})
\end{equation}
where, E{$_\textrm{ads}$} is the adsorption energy (see Table~\ref{Table4}) of the H$_2$O  on the vdW HTSs; E(adsorbed H$_\textrm{2}$O) is the energy of the vdW HTSs with adsorbed H$_2$O, E(vdW HTSs) is the energy of the vdW HTSs and E(H$_2$O) is the energy of the water molecule. From the Table~\ref{Table4} we observe, H{$_2$}O is physisorbed in the system and its binding strength in all the monolayers is higher than that in case of WS{$_2$}. Also, the transition metal oxides show higher binding strength than the transition metal dichalcogenides. The data supplement the analysis for considering the monolayers for Z-scheme photocatalysis.\\
\noindent In addition, we have calculated the Gibb's free energy change ($\Delta$G) of the intermediate in the HER as per the expression:
\begin{equation}
	\Delta \textrm{G} = \Delta \textrm{E}+\Delta \textrm{E}_\textrm{zpe} - \textrm{T}\Delta \textrm{S} 
\end{equation}
where $\Delta$E is the intermediate adsorption energy on the vdW HTSs, $\Delta$E$_\textrm{zpe}$ is the change in zero-point energy and $\Delta$S is the difference of entropy. The temperature, T, is taken as 300K. We have calculated the E$_\textrm{zpe}$ and TS by the following expression using Density Functional Perturbation Theory (DFPT):
\begin{equation}
	\textrm{E}{_\textrm{ZPE}} = \frac{1}{2}\sum_i \textrm{h}\nu_i
\end{equation}
\begin{equation}
	\textrm{TS} = \sum_i \textrm{h}\nu_i\frac{1}{\textrm{exp}(\frac{\textrm{h}\nu_i}{\textrm{k}_\textrm{B}\textrm{T}})-1} - \textrm{k}_\textrm{B}\textrm{T}\sum_i\textrm{ln} \left[1 - \textrm{exp}(-\frac{\textrm{h}\nu_i}{\textrm{k}_\textrm{B}\textrm{T}})\right]
\end{equation}

\noindent where h is Planck's constant, $\nu_i$ is vibrational frequencies and $\textrm{k}_\textrm{B}$ is Boltzmann constant. The HER diagram along the reaction pathway H$^+$ + e$^-$ $\,\to\,$ H* $\,\to\,$ 1/2 H{$_2$} is illustrated in the Fig. ~\ref{fig:4}. Here, H* represents the adsorbed intermediate. Under the consideration of pH = 0 and Standard Hydrogen Electrode (SHE) potential of 0 V the H$^+$ + e$^-$ is equivalent to the 1/2 H{$_2$}.
\\
\noindent Now, upon systematically revisiting all the aforementioned factors, we have observed vdW HTSs to be type II with their work function being affected by the interfacial potential difference. This further influences the charge separation. Another point is that the smaller conduction band offset between the two monolayer systems would promote efficient charge separation at the interface~\cite{khanchandani2014band}. Therefore, in reference to Fig. 3, with comparatively large difference in two conduction band levels along with type II alignment, the consideration for interlayer recombination in vdW HTSs for Z-scheme process seems very much plausible. Effective mass being the important factor for the same, we have indicated the recombination probability by the effective mass ratio (D). However, since this approach is qualitative we have explicitly included the results of optical response, H{$_2$}O adsorption and HER as direct evidences to validate MoSSe/HfS{$_2$}, MoSSe/TiS{$_2$}, MoS{$_2$}/T-SnO{$_2$}, MoSSe/T-SnO{$_2$}, MoS{$_2$}/ZrS{$_2$} and MoSSe/ZrS{$_2$} are probable Z-scheme photocatalysts.


\section{Conclusion}
\noindent An exhaustive study has been undertaken for understanding different MoS{$_2$} and MoSSe based vdW HTSs. As per the band edge alignment, it is observed that these vdW HTSs do not facilitate the normal photocatalytic process as they do not straddle the redox potential. However, these are predicted to be implementing the natural photocatalysis by Z-scheme. The band edge alignment of the MoSSe vdW HTSs has inferred large band gap in configuration III than in case of configuration II. This has been attributed to the additional anionic potential gradient due to Se atomic layer at the interface in configuration II. The probability of recombination as implicated by the effective mass ratio (D) has corroborated the MoSSe/HfS{$_2$}, MoSSe/TiS{$_2$}, MoS{$_2$}/T-SnO{$_2$}, MoSSe/T-SnO{$_2$}, MoS{$_2$}/ZrS{$_2$} and MoSSe/ZrS{$_2$} as favourable Z-scheme photocatalyst. The absorption spectra further confirm the visible light response of these vdW HTSs and hence, their applicability in the photocatalytic devices. Finally, the H{$_2$}O adsorption  and HER indicate their interaction with water to help in the photocatalytic process.\\


\section{Supporting Information}
 (I) Electrostatic Potential. (II) Lattice Parameters. (III) Band Gaps of Configurations. (IV) Bandstructures of Z-scheme vdW HTSs. (V) Planar Averaged Charged Density. (VI) Absorption Spectra of Monolayers. (VII) Exciton Binding Energy (VIII) Carrier Mobility.
 
 
\section{Acknowledgement}
\noindent AS acknowledges IIT Delhi for the financial support. AS and MJ thank Pooja Basera for helpful discussions. MJ acknowledges CSIR, India, for the senior research fellowship [grant no. 09/086(1344)/2018-EMR-I]. SB acknowledges the financial support SERB under his Core Research Grant [CRG/2019/000647]. We acknowledge the High Performance Computing (HPC) facility at IIT Delhi for computational resources.
\bibliography{ref}

\providecommand{\latin}[1]{#1}
\makeatletter
\providecommand{\doi}
  {\begingroup\let\do\@makeother\dospecials
  \catcode`\{=1 \catcode`\}=2 \doi@aux}
\providecommand{\doi@aux}[1]{\endgroup\texttt{#1}}
\makeatother
\providecommand*\mcitethebibliography{\thebibliography}
\csname @ifundefined\endcsname{endmcitethebibliography}
  {\let\endmcitethebibliography\endthebibliography}{}
\begin{mcitethebibliography}{83}
\providecommand*\natexlab[1]{#1}
\providecommand*\mciteSetBstSublistMode[1]{}
\providecommand*\mciteSetBstMaxWidthForm[2]{}
\providecommand*\mciteBstWouldAddEndPuncttrue
  {\def\EndOfBibitem{\unskip.}}
\providecommand*\mciteBstWouldAddEndPunctfalse
  {\let\EndOfBibitem\relax}
\providecommand*\mciteSetBstMidEndSepPunct[3]{}
\providecommand*\mciteSetBstSublistLabelBeginEnd[3]{}
\providecommand*\EndOfBibitem{}
\mciteSetBstSublistMode{f}
\mciteSetBstMaxWidthForm{subitem}{(\alph{mcitesubitemcount})}
\mciteSetBstSublistLabelBeginEnd
  {\mcitemaxwidthsubitemform\space}
  {\relax}
  {\relax}

\bibitem[Ni \latin{et~al.}(2007)Ni, Leung, Leung, and Sumathy]{ni2007review}
Ni,~M.; Leung,~M.~K.; Leung,~D.~Y.; Sumathy,~K. A review and recent
  developments in photocatalytic water-splitting using TiO2 for hydrogen
  production. \emph{Renewable and Sustainable Energy Reviews} \textbf{2007},
  \emph{11}, 401--425\relax
\mciteBstWouldAddEndPuncttrue
\mciteSetBstMidEndSepPunct{\mcitedefaultmidpunct}
{\mcitedefaultendpunct}{\mcitedefaultseppunct}\relax
\EndOfBibitem
\bibitem[Yao \latin{et~al.}(2015)Yao, Fan, Wang, Luo, and
  Fei]{yao2015pollutant}
Yao,~H.; Fan,~M.; Wang,~Y.; Luo,~G.; Fei,~W. Magnetic titanium dioxide based
  nanomaterials: synthesis, characteristics, and photocatalytic application in
  pollutant degradation. \emph{Journal of Materials Chemistry A} \textbf{2015},
  \emph{3}, 17511--17524\relax
\mciteBstWouldAddEndPuncttrue
\mciteSetBstMidEndSepPunct{\mcitedefaultmidpunct}
{\mcitedefaultendpunct}{\mcitedefaultseppunct}\relax
\EndOfBibitem
\bibitem[Opoku \latin{et~al.}(2017)Opoku, Govender, van Sittert, and
  Govender]{opoku2017role}
Opoku,~F.; Govender,~K.~K.; van Sittert,~C. G. C.~E.; Govender,~P.~P. Role of
  MoS 2 and WS 2 monolayers on photocatalytic hydrogen production and the
  pollutant degradation of monoclinic BiVO 4: A first-principles study.
  \emph{New Journal of Chemistry} \textbf{2017}, \emph{41}, 11701--11713\relax
\mciteBstWouldAddEndPuncttrue
\mciteSetBstMidEndSepPunct{\mcitedefaultmidpunct}
{\mcitedefaultendpunct}{\mcitedefaultseppunct}\relax
\EndOfBibitem
\bibitem[Yu \latin{et~al.}(2019)Yu, Chen, Zeng, Xie, Zhou, Liu, Wei, Yang, and
  Li]{yu2019facile}
Yu,~C.; Chen,~F.; Zeng,~D.; Xie,~Y.; Zhou,~W.; Liu,~Z.; Wei,~L.; Yang,~K.;
  Li,~D. A facile phase transformation strategy for fabrication of novel
  Z-scheme ternary heterojunctions with efficient photocatalytic properties.
  \emph{Nanoscale} \textbf{2019}, \emph{11}, 7720--7733\relax
\mciteBstWouldAddEndPuncttrue
\mciteSetBstMidEndSepPunct{\mcitedefaultmidpunct}
{\mcitedefaultendpunct}{\mcitedefaultseppunct}\relax
\EndOfBibitem
\bibitem[Gupta \latin{et~al.}(2015)Gupta, Sakthivel, and Seal]{gupta2015recent}
Gupta,~A.; Sakthivel,~T.; Seal,~S. Recent development in 2D materials beyond
  graphene. \emph{Progress in Materials Science} \textbf{2015}, \emph{73},
  44--126\relax
\mciteBstWouldAddEndPuncttrue
\mciteSetBstMidEndSepPunct{\mcitedefaultmidpunct}
{\mcitedefaultendpunct}{\mcitedefaultseppunct}\relax
\EndOfBibitem
\bibitem[Singh \latin{et~al.}(2020)Singh, Basera, Saini, Kumar, and
  Bhattacharya]{asjpcc}
Singh,~A.; Basera,~P.; Saini,~S.; Kumar,~M.; Bhattacharya,~S. Importance of
  Many-Body Dispersion in the Stability of Vacancies and Antisites in
  Free-Standing Monolayer of MoS2 from First-Principles Approaches. \emph{The
  Journal of Physical Chemistry C} \textbf{2020}, \emph{124}, 1390--1397\relax
\mciteBstWouldAddEndPuncttrue
\mciteSetBstMidEndSepPunct{\mcitedefaultmidpunct}
{\mcitedefaultendpunct}{\mcitedefaultseppunct}\relax
\EndOfBibitem
\bibitem[Chen \latin{et~al.}(2018)Chen, Huang, Ji, Adepalli, Yin, Ling, Wang,
  Xue, Dresselhaus, Kong, \latin{et~al.} others]{chen2018ACSNano}
Chen,~Y.; Huang,~S.; Ji,~X.; Adepalli,~K.; Yin,~K.; Ling,~X.; Wang,~X.;
  Xue,~J.; Dresselhaus,~M.; Kong,~J., \latin{et~al.}  Tuning electronic
  structure of single layer MoS2 through defect and interface engineering.
  \emph{ACS nano} \textbf{2018}, \emph{12}, 2569--2579\relax
\mciteBstWouldAddEndPuncttrue
\mciteSetBstMidEndSepPunct{\mcitedefaultmidpunct}
{\mcitedefaultendpunct}{\mcitedefaultseppunct}\relax
\EndOfBibitem
\bibitem[Rao \latin{et~al.}(2015)Rao, Gopalakrishnan, and
  Maitra]{rao2015comparative}
Rao,~C.; Gopalakrishnan,~K.; Maitra,~U. Comparative study of potential
  applications of graphene, MoS2, and other two-dimensional materials in energy
  devices, sensors, and related areas. \emph{ACS applied materials \&
  interfaces} \textbf{2015}, \emph{7}, 7809--7832\relax
\mciteBstWouldAddEndPuncttrue
\mciteSetBstMidEndSepPunct{\mcitedefaultmidpunct}
{\mcitedefaultendpunct}{\mcitedefaultseppunct}\relax
\EndOfBibitem
\bibitem[Singh \latin{et~al.}(2019)Singh, Singh, Kim, Yeom, and
  Nalwa]{singh2019ACSApplInter}
Singh,~E.; Singh,~P.; Kim,~K.~S.; Yeom,~G.~Y.; Nalwa,~H.~S. Flexible molybdenum
  disulfide (MoS2) atomic layers for wearable electronics and optoelectronics.
  \emph{ACS applied materials \& interfaces} \textbf{2019}, \emph{11},
  11061--11105\relax
\mciteBstWouldAddEndPuncttrue
\mciteSetBstMidEndSepPunct{\mcitedefaultmidpunct}
{\mcitedefaultendpunct}{\mcitedefaultseppunct}\relax
\EndOfBibitem
\bibitem[Zeng \latin{et~al.}(2013)Zeng, Liu, Dai, Yan, Zhu, He, Xie, Xu, Chen,
  Yao, \latin{et~al.} others]{zeng2013optical}
Zeng,~H.; Liu,~G.-B.; Dai,~J.; Yan,~Y.; Zhu,~B.; He,~R.; Xie,~L.; Xu,~S.;
  Chen,~X.; Yao,~W., \latin{et~al.}  Optical signature of symmetry variations
  and spin-valley coupling in atomically thin tungsten dichalcogenides.
  \emph{Scientific reports} \textbf{2013}, \emph{3}, 1608\relax
\mciteBstWouldAddEndPuncttrue
\mciteSetBstMidEndSepPunct{\mcitedefaultmidpunct}
{\mcitedefaultendpunct}{\mcitedefaultseppunct}\relax
\EndOfBibitem
\bibitem[Wang and Guo(2015)Wang, and Guo]{wang2015JPCC}
Wang,~C.-Y.; Guo,~G.-Y. Nonlinear optical properties of transition-metal
  dichalcogenide MX2 (M= Mo, W; X= S, Se) monolayers and trilayers from
  first-principles calculations. \emph{The Journal of Physical Chemistry C}
  \textbf{2015}, \emph{119}, 13268--13276\relax
\mciteBstWouldAddEndPuncttrue
\mciteSetBstMidEndSepPunct{\mcitedefaultmidpunct}
{\mcitedefaultendpunct}{\mcitedefaultseppunct}\relax
\EndOfBibitem
\bibitem[Glebko \latin{et~al.}(2018)Glebko, Aleksandrova, Tewari, Tripathi,
  Karppinen, and Karttunen]{glebko2018electronic}
Glebko,~N.; Aleksandrova,~I.; Tewari,~G.~C.; Tripathi,~T.~S.; Karppinen,~M.;
  Karttunen,~A.~J. Electronic and vibrational properties of TiS2, ZrS2, and
  HfS2: Periodic trends studied by dispersion-corrected hybrid density
  functional methods. \emph{The Journal of Physical Chemistry C} \textbf{2018},
  \emph{122}, 26835--26844\relax
\mciteBstWouldAddEndPuncttrue
\mciteSetBstMidEndSepPunct{\mcitedefaultmidpunct}
{\mcitedefaultendpunct}{\mcitedefaultseppunct}\relax
\EndOfBibitem
\bibitem[Lau \latin{et~al.}(2019)Lau, Cocchi, and Draxl]{lau2019electronic}
Lau,~K.~W.; Cocchi,~C.; Draxl,~C. Electronic and optical excitations of
  two-dimensional ZrS 2 and HfS 2 and their heterostructure. \emph{Physical
  Review Materials} \textbf{2019}, \emph{3}, 074001\relax
\mciteBstWouldAddEndPuncttrue
\mciteSetBstMidEndSepPunct{\mcitedefaultmidpunct}
{\mcitedefaultendpunct}{\mcitedefaultseppunct}\relax
\EndOfBibitem
\bibitem[Mattinen \latin{et~al.}(2019)Mattinen, Popov, Vehkamäki, King,
  Mizohata, Jalkanen, Räisänen, Leskelä, and Ritala]{mattinen2019atomic}
Mattinen,~M.; Popov,~G.; Vehkamäki,~M.; King,~P.~J.; Mizohata,~K.;
  Jalkanen,~P.; Räisänen,~J.; Leskelä,~M.; Ritala,~M. Atomic layer
  deposition of emerging 2D semiconductors, HfS2 and ZrS2, for optoelectronics.
  \emph{Chemistry of Materials} \textbf{2019}, \emph{31}, 5713--5724\relax
\mciteBstWouldAddEndPuncttrue
\mciteSetBstMidEndSepPunct{\mcitedefaultmidpunct}
{\mcitedefaultendpunct}{\mcitedefaultseppunct}\relax
\EndOfBibitem
\bibitem[Leb{\`e}gue \latin{et~al.}(2013)Leb{\`e}gue, Bj{\"o}rkman,
  Klintenberg, Nieminen, and Eriksson]{lebegue2013two}
Leb{\`e}gue,~S.; Bj{\"o}rkman,~T.; Klintenberg,~M.; Nieminen,~R.~M.;
  Eriksson,~O. Two-dimensional materials from data filtering and ab initio
  calculations. \emph{Physical Review X} \textbf{2013}, \emph{3}, 031002\relax
\mciteBstWouldAddEndPuncttrue
\mciteSetBstMidEndSepPunct{\mcitedefaultmidpunct}
{\mcitedefaultendpunct}{\mcitedefaultseppunct}\relax
\EndOfBibitem
\bibitem[Yang \latin{et~al.}(2019)Yang, Zeng, Kang, Betzler, Czarnik, Zhang,
  Ophus, Yu, Bustillo, Pan, \latin{et~al.} others]{yang2019formation}
Yang,~J.; Zeng,~Z.; Kang,~J.; Betzler,~S.; Czarnik,~C.; Zhang,~X.; Ophus,~C.;
  Yu,~C.; Bustillo,~K.; Pan,~M., \latin{et~al.}  Formation of two-dimensional
  transition metal oxide nanosheets with nanoparticles as intermediates.
  \emph{Nature materials} \textbf{2019}, \emph{18}, 970--976\relax
\mciteBstWouldAddEndPuncttrue
\mciteSetBstMidEndSepPunct{\mcitedefaultmidpunct}
{\mcitedefaultendpunct}{\mcitedefaultseppunct}\relax
\EndOfBibitem
\bibitem[Leong \latin{et~al.}(2016)Leong, Pan, and Ho]{leong2016two}
Leong,~C.~C.; Pan,~H.; Ho,~S.~K. Two-dimensional transition-metal oxide
  monolayers as cathode materials for Li and Na ion batteries. \emph{Physical
  Chemistry Chemical Physics} \textbf{2016}, \emph{18}, 7527--7534\relax
\mciteBstWouldAddEndPuncttrue
\mciteSetBstMidEndSepPunct{\mcitedefaultmidpunct}
{\mcitedefaultendpunct}{\mcitedefaultseppunct}\relax
\EndOfBibitem
\bibitem[Liao and Carter(2013)Liao, and Carter]{liao2013new}
Liao,~P.; Carter,~E.~A. New concepts and modeling strategies to design and
  evaluate photo-electro-catalysts based on transition metal oxides.
  \emph{Chemical Society Reviews} \textbf{2013}, \emph{42}, 2401--2422\relax
\mciteBstWouldAddEndPuncttrue
\mciteSetBstMidEndSepPunct{\mcitedefaultmidpunct}
{\mcitedefaultendpunct}{\mcitedefaultseppunct}\relax
\EndOfBibitem
\bibitem[Ju \latin{et~al.}(2020)Ju, Bie, Shang, Tang, and Kou]{ju2020janus}
Ju,~L.; Bie,~M.; Shang,~J.; Tang,~X.; Kou,~L. Janus transition metal
  dichalcogenides: A superior platform for photocatalytic water splitting.
  \emph{Journal of Physics: Materials} \textbf{2020}, \emph{3}, 022004\relax
\mciteBstWouldAddEndPuncttrue
\mciteSetBstMidEndSepPunct{\mcitedefaultmidpunct}
{\mcitedefaultendpunct}{\mcitedefaultseppunct}\relax
\EndOfBibitem
\bibitem[Sun \latin{et~al.}(2018)Sun, Shuai, and Wang]{sun2018janus}
Sun,~Y.; Shuai,~Z.; Wang,~D. Janus monolayer of WSeTe, a new structural phase
  transition material driven by electrostatic gating. \emph{Nanoscale}
  \textbf{2018}, \emph{10}, 21629--21633\relax
\mciteBstWouldAddEndPuncttrue
\mciteSetBstMidEndSepPunct{\mcitedefaultmidpunct}
{\mcitedefaultendpunct}{\mcitedefaultseppunct}\relax
\EndOfBibitem
\bibitem[Song \latin{et~al.}(2019)Song, Liu, and Yam]{song2019suppressed}
Song,~B.; Liu,~L.; Yam,~C. Suppressed Carrier Recombination in Janus MoSSe
  Bilayer Stacks: A Time-Domain Ab Initio Study. \emph{The journal of physical
  chemistry letters} \textbf{2019}, \emph{10}, 5564--5570\relax
\mciteBstWouldAddEndPuncttrue
\mciteSetBstMidEndSepPunct{\mcitedefaultmidpunct}
{\mcitedefaultendpunct}{\mcitedefaultseppunct}\relax
\EndOfBibitem
\bibitem[Sun and Schwingenschlogl(2020)Sun, and Schwingenschlogl]{sun2020b2p6}
Sun,~M.; Schwingenschlogl,~U. B2P6: A Two-dimensional anisotropic Janus
  material with potential in photocatalytic water splitting and metal-ion
  batteries. \emph{Chemistry of Materials} \textbf{2020}, \emph{32},
  4795--4800\relax
\mciteBstWouldAddEndPuncttrue
\mciteSetBstMidEndSepPunct{\mcitedefaultmidpunct}
{\mcitedefaultendpunct}{\mcitedefaultseppunct}\relax
\EndOfBibitem
\bibitem[Ma \latin{et~al.}(2018)Ma, Wu, Wang, and Wang]{ma2018janus}
Ma,~X.; Wu,~X.; Wang,~H.; Wang,~Y. A Janus MoSSe monolayer: a potential wide
  solar-spectrum water-splitting photocatalyst with a low carrier recombination
  rate. \emph{Journal of Materials Chemistry A} \textbf{2018}, \emph{6},
  2295--2301\relax
\mciteBstWouldAddEndPuncttrue
\mciteSetBstMidEndSepPunct{\mcitedefaultmidpunct}
{\mcitedefaultendpunct}{\mcitedefaultseppunct}\relax
\EndOfBibitem
\bibitem[Luo \latin{et~al.}(2020)Luo, Wang, Shu, Chou, Ren, Yu, and
  Sun]{luo2020mosse}
Luo,~Y.; Wang,~S.; Shu,~H.; Chou,~J.-P.; Ren,~K.; Yu,~J.; Sun,~M. A MoSSe/blue
  phosphorene vdw heterostructure with energy conversion efficiency of 19.9\%
  for photocatalytic water splitting. \emph{Semiconductor Science and
  Technology} \textbf{2020}, \emph{35}, 125008\relax
\mciteBstWouldAddEndPuncttrue
\mciteSetBstMidEndSepPunct{\mcitedefaultmidpunct}
{\mcitedefaultendpunct}{\mcitedefaultseppunct}\relax
\EndOfBibitem
\bibitem[Kuc and Heine(2015)Kuc, and Heine]{kuc2015ChemSocRev}
Kuc,~A.; Heine,~T. The electronic structure calculations of two-dimensional
  transition-metal dichalcogenides in the presence of external electric and
  magnetic fields. \emph{Chemical Society Reviews} \textbf{2015}, \emph{44},
  2603--2614\relax
\mciteBstWouldAddEndPuncttrue
\mciteSetBstMidEndSepPunct{\mcitedefaultmidpunct}
{\mcitedefaultendpunct}{\mcitedefaultseppunct}\relax
\EndOfBibitem
\bibitem[Kumar \latin{et~al.}(2018)Kumar, Verzhbitskiy, Giustiniano,
  Sidiropoulos, Oulton, and Eda]{kumar2018interlayer}
Kumar,~R.; Verzhbitskiy,~I.; Giustiniano,~F.; Sidiropoulos,~T.~P.;
  Oulton,~R.~F.; Eda,~G. Interlayer screening effects in WS2/WSe2 van der Waals
  hetero-bilayer. \emph{2D Materials} \textbf{2018}, \emph{5}, 041003\relax
\mciteBstWouldAddEndPuncttrue
\mciteSetBstMidEndSepPunct{\mcitedefaultmidpunct}
{\mcitedefaultendpunct}{\mcitedefaultseppunct}\relax
\EndOfBibitem
\bibitem[Huang \latin{et~al.}(2017)Huang, Chen, Wang, Peng, Qian, and
  Wang]{huang2017layer}
Huang,~Y.; Chen,~X.; Wang,~C.; Peng,~L.; Qian,~Q.; Wang,~S. Layer-dependent
  electronic properties of phosphorene-like materials and phosphorene-based van
  der Waals heterostructures. \emph{Nanoscale} \textbf{2017}, \emph{9},
  8616--8622\relax
\mciteBstWouldAddEndPuncttrue
\mciteSetBstMidEndSepPunct{\mcitedefaultmidpunct}
{\mcitedefaultendpunct}{\mcitedefaultseppunct}\relax
\EndOfBibitem
\bibitem[Terrones \latin{et~al.}(2013)Terrones, L{\'o}pez-Ur{\'\i}as, and
  Terrones]{terrones2013novel}
Terrones,~H.; L{\'o}pez-Ur{\'\i}as,~F.; Terrones,~M. Novel hetero-layered
  materials with tunable direct band gaps by sandwiching different metal
  disulfides and diselenides. \emph{Scientific reports} \textbf{2013},
  \emph{3}, 1--7\relax
\mciteBstWouldAddEndPuncttrue
\mciteSetBstMidEndSepPunct{\mcitedefaultmidpunct}
{\mcitedefaultendpunct}{\mcitedefaultseppunct}\relax
\EndOfBibitem
\bibitem[Zhang and Schwingenschl{\"o}gl(2018)Zhang, and
  Schwingenschl{\"o}gl]{zhang2018rashba}
Zhang,~Q.; Schwingenschl{\"o}gl,~U. Rashba effect and enriched spin-valley
  coupling in Ga X/M X 2 (M= Mo, W; X= S, Se, Te) heterostructures.
  \emph{Physical Review B} \textbf{2018}, \emph{97}, 155415\relax
\mciteBstWouldAddEndPuncttrue
\mciteSetBstMidEndSepPunct{\mcitedefaultmidpunct}
{\mcitedefaultendpunct}{\mcitedefaultseppunct}\relax
\EndOfBibitem
\bibitem[Rivera \latin{et~al.}(2016)Rivera, Seyler, Yu, Schaibley, Yan,
  Mandrus, Yao, and Xu]{rivera2016valley}
Rivera,~P.; Seyler,~K.~L.; Yu,~H.; Schaibley,~J.~R.; Yan,~J.; Mandrus,~D.~G.;
  Yao,~W.; Xu,~X. Valley-polarized exciton dynamics in a 2D semiconductor
  heterostructure. \emph{Science} \textbf{2016}, \emph{351}, 688--691\relax
\mciteBstWouldAddEndPuncttrue
\mciteSetBstMidEndSepPunct{\mcitedefaultmidpunct}
{\mcitedefaultendpunct}{\mcitedefaultseppunct}\relax
\EndOfBibitem
\bibitem[Li \latin{et~al.}(2018)Li, Zhang, Li, Liang, and
  Zeng]{li2018multifunctional}
Li,~P.; Zhang,~W.; Li,~D.; Liang,~C.; Zeng,~X.~C. Multifunctional Binary
  Monolayers Ge x P y: Tunable Band Gap, Ferromagnetism, and Photocatalyst for
  Water Splitting. \emph{ACS applied materials \& interfaces} \textbf{2018},
  \emph{10}, 19897--19905\relax
\mciteBstWouldAddEndPuncttrue
\mciteSetBstMidEndSepPunct{\mcitedefaultmidpunct}
{\mcitedefaultendpunct}{\mcitedefaultseppunct}\relax
\EndOfBibitem
\bibitem[Li \latin{et~al.}(2019)Li, Zhang, Liang, and Zeng]{li2019two}
Li,~P.; Zhang,~W.; Liang,~C.; Zeng,~X.~C. Two-dimensional MgX 2 Se 4 (X= Al,
  Ga) monolayers with tunable electronic properties for optoelectronic and
  photocatalytic applications. \emph{Nanoscale} \textbf{2019}, \emph{11},
  19806--19813\relax
\mciteBstWouldAddEndPuncttrue
\mciteSetBstMidEndSepPunct{\mcitedefaultmidpunct}
{\mcitedefaultendpunct}{\mcitedefaultseppunct}\relax
\EndOfBibitem
\bibitem[Luo \latin{et~al.}(2019)Luo, Ren, Wang, Chou, Yu, Sun, and
  Sun]{luo2019first}
Luo,~Y.; Ren,~K.; Wang,~S.; Chou,~J.-P.; Yu,~J.; Sun,~Z.; Sun,~M.
  First-principles study on transition-metal dichalcogenide/BSe van der Waals
  heterostructures: A promising water-splitting photocatalyst. \emph{The
  Journal of Physical Chemistry C} \textbf{2019}, \emph{123},
  22742--22751\relax
\mciteBstWouldAddEndPuncttrue
\mciteSetBstMidEndSepPunct{\mcitedefaultmidpunct}
{\mcitedefaultendpunct}{\mcitedefaultseppunct}\relax
\EndOfBibitem
\bibitem[Luo \latin{et~al.}(2019)Luo, Wang, Ren, Chou, Yu, Sun, and
  Sun]{luo2019transition}
Luo,~Y.; Wang,~S.; Ren,~K.; Chou,~J.-P.; Yu,~J.; Sun,~Z.; Sun,~M.
  Transition-metal dichalcogenides/Mg (OH) 2 van der Waals heterostructures as
  promising water-splitting photocatalysts: a first-principles study.
  \emph{Physical Chemistry Chemical Physics} \textbf{2019}, \emph{21},
  1791--1796\relax
\mciteBstWouldAddEndPuncttrue
\mciteSetBstMidEndSepPunct{\mcitedefaultmidpunct}
{\mcitedefaultendpunct}{\mcitedefaultseppunct}\relax
\EndOfBibitem
\bibitem[Ren \latin{et~al.}(2020)Ren, Wang, Luo, Chou, Yu, Tang, and
  Sun]{ren2020high}
Ren,~K.; Wang,~S.; Luo,~Y.; Chou,~J.-P.; Yu,~J.; Tang,~W.; Sun,~M.
  High-efficiency photocatalyst for water splitting: a Janus MoSSe/XN (X= Ga,
  Al) van der Waals heterostructure. \emph{Journal of Physics D: Applied
  Physics} \textbf{2020}, \emph{53}, 185504\relax
\mciteBstWouldAddEndPuncttrue
\mciteSetBstMidEndSepPunct{\mcitedefaultmidpunct}
{\mcitedefaultendpunct}{\mcitedefaultseppunct}\relax
\EndOfBibitem
\bibitem[Wang \latin{et~al.}(2018)Wang, Ren, Tian, Yu, and Sun]{wang2018mos}
Wang,~S.; Ren,~C.; Tian,~H.; Yu,~J.; Sun,~M. MoS 2/ZnO van der Waals
  heterostructure as a high-efficiency water splitting photocatalyst: a
  first-principles study. \emph{Physical Chemistry Chemical Physics}
  \textbf{2018}, \emph{20}, 13394--13399\relax
\mciteBstWouldAddEndPuncttrue
\mciteSetBstMidEndSepPunct{\mcitedefaultmidpunct}
{\mcitedefaultendpunct}{\mcitedefaultseppunct}\relax
\EndOfBibitem
\bibitem[Li \latin{et~al.}(2013)Li, Li, Araujo, Luo, and Ahuja]{li2013single}
Li,~Y.; Li,~Y.-L.; Araujo,~C.~M.; Luo,~W.; Ahuja,~R. Single-layer MoS 2 as an
  efficient photocatalyst. \emph{Catalysis Science \& Technology}
  \textbf{2013}, \emph{3}, 2214--2220\relax
\mciteBstWouldAddEndPuncttrue
\mciteSetBstMidEndSepPunct{\mcitedefaultmidpunct}
{\mcitedefaultendpunct}{\mcitedefaultseppunct}\relax
\EndOfBibitem
\bibitem[Parzinger \latin{et~al.}(2015)Parzinger, Miller, Blaschke, Garrido,
  Ager, Holleitner, and Wurstbauer]{parzinger2015photocatalytic}
Parzinger,~E.; Miller,~B.; Blaschke,~B.; Garrido,~J.~A.; Ager,~J.~W.;
  Holleitner,~A.; Wurstbauer,~U. Photocatalytic stability of single-and
  few-layer MoS2. \emph{ACS nano} \textbf{2015}, \emph{9}, 11302--11309\relax
\mciteBstWouldAddEndPuncttrue
\mciteSetBstMidEndSepPunct{\mcitedefaultmidpunct}
{\mcitedefaultendpunct}{\mcitedefaultseppunct}\relax
\EndOfBibitem
\bibitem[Tang \latin{et~al.}(2016)Tang, Yin, Zhang, Wen, Zhang, Liu, and
  Lau]{tang2016spatial}
Tang,~Z.-K.; Yin,~W.-J.; Zhang,~L.; Wen,~B.; Zhang,~D.-Y.; Liu,~L.-M.;
  Lau,~W.-M. Spatial separation of photo-generated electron-hole pairs in
  BiOBr/BiOI bilayer to facilitate water splitting. \emph{Scientific reports}
  \textbf{2016}, \emph{6}, 32764\relax
\mciteBstWouldAddEndPuncttrue
\mciteSetBstMidEndSepPunct{\mcitedefaultmidpunct}
{\mcitedefaultendpunct}{\mcitedefaultseppunct}\relax
\EndOfBibitem
\bibitem[Gao \latin{et~al.}(2019)Gao, Shen, Ma, Wu, and Zhou]{gao2019water}
Gao,~X.; Shen,~Y.; Ma,~Y.; Wu,~S.; Zhou,~Z. A water splitting photocatalysis:
  blue phosphorus/g-GeC van der Waals heterostructure. \emph{Applied Physics
  Letters} \textbf{2019}, \emph{114}, 093902\relax
\mciteBstWouldAddEndPuncttrue
\mciteSetBstMidEndSepPunct{\mcitedefaultmidpunct}
{\mcitedefaultendpunct}{\mcitedefaultseppunct}\relax
\EndOfBibitem
\bibitem[Zhang \latin{et~al.}(2018)Zhang, Zhang, Zheng, Gao, Zhao, and
  Yang]{Direct_ZScheme_Zhang}
Zhang,~R.; Zhang,~L.; Zheng,~Q.; Gao,~P.; Zhao,~J.; Yang,~J. Direct Z-scheme
  water splitting photocatalyst based on two-dimensional Van Der Waals
  heterostructures. \emph{The journal of physical chemistry letters}
  \textbf{2018}, \emph{9}, 5419--5424\relax
\mciteBstWouldAddEndPuncttrue
\mciteSetBstMidEndSepPunct{\mcitedefaultmidpunct}
{\mcitedefaultendpunct}{\mcitedefaultseppunct}\relax
\EndOfBibitem
\bibitem[Ren \latin{et~al.}(2019)Ren, Ren, Luo, Xu, Yu, Tang, and
  Sun]{ren2019using}
Ren,~K.; Ren,~C.; Luo,~Y.; Xu,~Y.; Yu,~J.; Tang,~W.; Sun,~M. Using van der
  Waals heterostructures based on two-dimensional blue phosphorus and XC (X=
  Ge, Si) for water-splitting photocatalysis: a first-principles study.
  \emph{Physical Chemistry Chemical Physics} \textbf{2019}, \emph{21},
  9949--9956\relax
\mciteBstWouldAddEndPuncttrue
\mciteSetBstMidEndSepPunct{\mcitedefaultmidpunct}
{\mcitedefaultendpunct}{\mcitedefaultseppunct}\relax
\EndOfBibitem
\bibitem[Wang \latin{et~al.}(2018)Wang, Li, Zhao, Cai, Yu, Li, Liu, Zhang, and
  Ke]{wang2018electronic}
Wang,~B.-J.; Li,~X.-H.; Zhao,~R.; Cai,~X.-L.; Yu,~W.-Y.; Li,~W.-B.; Liu,~Z.-S.;
  Zhang,~L.-W.; Ke,~S.-H. Electronic structures and enhanced photocatalytic
  properties of blue phosphorene/BSe van der Waals heterostructures.
  \emph{Journal of Materials Chemistry A} \textbf{2018}, \emph{6},
  8923--8929\relax
\mciteBstWouldAddEndPuncttrue
\mciteSetBstMidEndSepPunct{\mcitedefaultmidpunct}
{\mcitedefaultendpunct}{\mcitedefaultseppunct}\relax
\EndOfBibitem
\bibitem[Fu \latin{et~al.}(2016)Fu, Luo, Li, and Yang]{Direct_Z_Fu}
Fu,~C.-F.; Luo,~Q.; Li,~X.; Yang,~J. Two-dimensional van der Waals
  nanocomposites as Z-scheme type photocatalysts for hydrogen production from
  overall water splitting. \emph{Journal of Materials Chemistry A}
  \textbf{2016}, \emph{4}, 18892--18898\relax
\mciteBstWouldAddEndPuncttrue
\mciteSetBstMidEndSepPunct{\mcitedefaultmidpunct}
{\mcitedefaultendpunct}{\mcitedefaultseppunct}\relax
\EndOfBibitem
\bibitem[Liu \latin{et~al.}(2020)Liu, Zeng, Easton, Li, Xia, Yin, Hu, Hu, Xia,
  McCarthy, \latin{et~al.} others]{liu2020situ}
Liu,~Y.; Zeng,~X.; Easton,~C.~D.; Li,~Q.; Xia,~Y.; Yin,~Y.; Hu,~X.; Hu,~J.;
  Xia,~D.; McCarthy,~D.~T., \latin{et~al.}  An in situ assembled WO 3--TiO 2
  vertical heterojunction for enhanced Z-scheme photocatalytic activity.
  \emph{Nanoscale} \textbf{2020}, \emph{12}, 8775--8784\relax
\mciteBstWouldAddEndPuncttrue
\mciteSetBstMidEndSepPunct{\mcitedefaultmidpunct}
{\mcitedefaultendpunct}{\mcitedefaultseppunct}\relax
\EndOfBibitem
\bibitem[Xia \latin{et~al.}(2019)Xia, Song, Wang, Zhang, Sui, Zhang, Colvin,
  and William]{xia2019latest}
Xia,~X.; Song,~M.; Wang,~H.; Zhang,~X.; Sui,~N.; Zhang,~Q.; Colvin,~V.~L.;
  William,~W.~Y. Latest progress in constructing solid-state Z scheme
  photocatalysts for water splitting. \emph{Nanoscale} \textbf{2019},
  \emph{11}, 11071--11082\relax
\mciteBstWouldAddEndPuncttrue
\mciteSetBstMidEndSepPunct{\mcitedefaultmidpunct}
{\mcitedefaultendpunct}{\mcitedefaultseppunct}\relax
\EndOfBibitem
\bibitem[Maeda(2013)]{maeda2013z}
Maeda,~K. Z-scheme water splitting using two different semiconductor
  photocatalysts. \emph{ACS Catalysis} \textbf{2013}, \emph{3},
  1486--1503\relax
\mciteBstWouldAddEndPuncttrue
\mciteSetBstMidEndSepPunct{\mcitedefaultmidpunct}
{\mcitedefaultendpunct}{\mcitedefaultseppunct}\relax
\EndOfBibitem
\bibitem[Li \latin{et~al.}(2016)Li, Tu, Zhou, and Zou]{li2016z}
Li,~H.; Tu,~W.; Zhou,~Y.; Zou,~Z. Z-Scheme photocatalytic systems for promoting
  photocatalytic performance: recent progress and future challenges.
  \emph{Advanced Science} \textbf{2016}, \emph{3}, 1500389\relax
\mciteBstWouldAddEndPuncttrue
\mciteSetBstMidEndSepPunct{\mcitedefaultmidpunct}
{\mcitedefaultendpunct}{\mcitedefaultseppunct}\relax
\EndOfBibitem
\bibitem[Bard(1979)]{bard1979photoelectrochemistry}
Bard,~A.~J. Photoelectrochemistry and heterogeneous photocatalysis at
  semiconductors. \emph{Journal of Photochemistry} \textbf{1979}, \emph{10},
  59--75\relax
\mciteBstWouldAddEndPuncttrue
\mciteSetBstMidEndSepPunct{\mcitedefaultmidpunct}
{\mcitedefaultendpunct}{\mcitedefaultseppunct}\relax
\EndOfBibitem
\bibitem[Lang and Hu(2020)Lang, and Hu]{lang2020phosphorus}
Lang,~J.; Hu,~Y.~H. Phosphorus-based metal-free Z-scheme 2D van der Waals
  heterostructures for visible-light photocatalytic water splitting: a
  first-principles study. \emph{Physical Chemistry Chemical Physics}
  \textbf{2020}, \emph{22}, 9250--9256\relax
\mciteBstWouldAddEndPuncttrue
\mciteSetBstMidEndSepPunct{\mcitedefaultmidpunct}
{\mcitedefaultendpunct}{\mcitedefaultseppunct}\relax
\EndOfBibitem
\bibitem[Ju \latin{et~al.}(2018)Ju, Dai, Wei, Li, and Huang]{ju2018dft}
Ju,~L.; Dai,~Y.; Wei,~W.; Li,~M.; Huang,~B. DFT investigation on
  two-dimensional GeS/WS2 van der Waals heterostructure for direct Z-scheme
  photocatalytic overall water splitting. \emph{Applied Surface Science}
  \textbf{2018}, \emph{434}, 365--374\relax
\mciteBstWouldAddEndPuncttrue
\mciteSetBstMidEndSepPunct{\mcitedefaultmidpunct}
{\mcitedefaultendpunct}{\mcitedefaultseppunct}\relax
\EndOfBibitem
\bibitem[Rasmussen and Thygesen(2015)Rasmussen, and
  Thygesen]{rasmussen2015computational}
Rasmussen,~F.~A.; Thygesen,~K.~S. Computational 2D materials database:
  electronic structure of transition-metal dichalcogenides and oxides.
  \emph{The Journal of Physical Chemistry C} \textbf{2015}, \emph{119},
  13169--13183\relax
\mciteBstWouldAddEndPuncttrue
\mciteSetBstMidEndSepPunct{\mcitedefaultmidpunct}
{\mcitedefaultendpunct}{\mcitedefaultseppunct}\relax
\EndOfBibitem
\bibitem[Martin(2004)]{martin2004electronic}
Martin,~R.~M. \emph{Electronic structure: basic theory and practical methods};
  Cambridge university press, 2004\relax
\mciteBstWouldAddEndPuncttrue
\mciteSetBstMidEndSepPunct{\mcitedefaultmidpunct}
{\mcitedefaultendpunct}{\mcitedefaultseppunct}\relax
\EndOfBibitem
\bibitem[Martin \latin{et~al.}(2016)Martin, Reining, and
  Ceperley]{martin2016interacting}
Martin,~R.~M.; Reining,~L.; Ceperley,~D.~M. \emph{Interacting Electrons};
  Cambridge University Press, 2016\relax
\mciteBstWouldAddEndPuncttrue
\mciteSetBstMidEndSepPunct{\mcitedefaultmidpunct}
{\mcitedefaultendpunct}{\mcitedefaultseppunct}\relax
\EndOfBibitem
\bibitem[Freysoldt \latin{et~al.}(2014)Freysoldt, Grabowski, Hickel,
  Neugebauer, Kresse, Janotti, and Van~de Walle]{freysoldt2014RevModPhys}
Freysoldt,~C.; Grabowski,~B.; Hickel,~T.; Neugebauer,~J.; Kresse,~G.;
  Janotti,~A.; Van~de Walle,~C.~G. First-principles calculations for point
  defects in solids. \emph{Reviews of modern physics} \textbf{2014}, \emph{86},
  253\relax
\mciteBstWouldAddEndPuncttrue
\mciteSetBstMidEndSepPunct{\mcitedefaultmidpunct}
{\mcitedefaultendpunct}{\mcitedefaultseppunct}\relax
\EndOfBibitem
\bibitem[Feng \latin{et~al.}(2014)Feng, Su, Chen, and
  Liu]{feng2014MaterChemPhys}
Feng,~L.-p.; Su,~J.; Chen,~S.; Liu,~Z.-t. First-principles investigations on
  vacancy formation and electronic structures of monolayer MoS2.
  \emph{Materials Chemistry and Physics} \textbf{2014}, \emph{148}, 5--9\relax
\mciteBstWouldAddEndPuncttrue
\mciteSetBstMidEndSepPunct{\mcitedefaultmidpunct}
{\mcitedefaultendpunct}{\mcitedefaultseppunct}\relax
\EndOfBibitem
\bibitem[Hohenberg and Kohn(1964)Hohenberg, and Kohn]{hohenberg1964PhysRev}
Hohenberg,~P.; Kohn,~W. Inhomogeneous electron gas. \emph{Physical review}
  \textbf{1964}, \emph{136}, B864\relax
\mciteBstWouldAddEndPuncttrue
\mciteSetBstMidEndSepPunct{\mcitedefaultmidpunct}
{\mcitedefaultendpunct}{\mcitedefaultseppunct}\relax
\EndOfBibitem
\bibitem[Kohn and Sham(1965)Kohn, and Sham]{kohn1965PhysRev}
Kohn,~W.; Sham,~L.~J. Self-consistent equations including exchange and
  correlation effects. \emph{Physical review} \textbf{1965}, \emph{140},
  A1133\relax
\mciteBstWouldAddEndPuncttrue
\mciteSetBstMidEndSepPunct{\mcitedefaultmidpunct}
{\mcitedefaultendpunct}{\mcitedefaultseppunct}\relax
\EndOfBibitem
\bibitem[Kresse and Furthm{\"u}ller(1996)Kresse, and
  Furthm{\"u}ller]{kresse1996efficient}
Kresse,~G.; Furthm{\"u}ller,~J. Efficient iterative schemes for ab initio
  total-energy calculations using a plane-wave basis set. \emph{Physical review
  B} \textbf{1996}, \emph{54}, 11169\relax
\mciteBstWouldAddEndPuncttrue
\mciteSetBstMidEndSepPunct{\mcitedefaultmidpunct}
{\mcitedefaultendpunct}{\mcitedefaultseppunct}\relax
\EndOfBibitem
\bibitem[Bl{\"o}chl(1994)]{blochl1994projector}
Bl{\"o}chl,~P.~E. Projector augmented-wave method. \emph{Physical review B}
  \textbf{1994}, \emph{50}, 17953\relax
\mciteBstWouldAddEndPuncttrue
\mciteSetBstMidEndSepPunct{\mcitedefaultmidpunct}
{\mcitedefaultendpunct}{\mcitedefaultseppunct}\relax
\EndOfBibitem
\bibitem[Blum \latin{et~al.}(2009)Blum, Gehrke, Hanke, Havu, Havu, Ren, Reuter,
  and Scheffler]{blum2009ComputPhysCommun}
Blum,~V.; Gehrke,~R.; Hanke,~F.; Havu,~P.; Havu,~V.; Ren,~X.; Reuter,~K.;
  Scheffler,~M. Ab initio molecular simulations with numeric atom-centered
  orbitals. \emph{Computer Physics Communications} \textbf{2009}, \emph{180},
  2175--2196\relax
\mciteBstWouldAddEndPuncttrue
\mciteSetBstMidEndSepPunct{\mcitedefaultmidpunct}
{\mcitedefaultendpunct}{\mcitedefaultseppunct}\relax
\EndOfBibitem
\bibitem[Stampfl and Van~de Walle(1999)Stampfl, and Van~de
  Walle]{stampfl1999PRB}
Stampfl,~C.; Van~de Walle,~C. Density-functional calculations for III-V
  nitrides using the local-density approximation and the generalized gradient
  approximation. \emph{Physical Review B} \textbf{1999}, \emph{59}, 5521\relax
\mciteBstWouldAddEndPuncttrue
\mciteSetBstMidEndSepPunct{\mcitedefaultmidpunct}
{\mcitedefaultendpunct}{\mcitedefaultseppunct}\relax
\EndOfBibitem
\bibitem[Perdew \latin{et~al.}(1996)Perdew, Burke, and
  Ernzerhof]{perdew1996PRL}
Perdew,~J.~P.; Burke,~K.; Ernzerhof,~M. Generalized gradient approximation made
  simple. \emph{Physical review letters} \textbf{1996}, \emph{77}, 3865\relax
\mciteBstWouldAddEndPuncttrue
\mciteSetBstMidEndSepPunct{\mcitedefaultmidpunct}
{\mcitedefaultendpunct}{\mcitedefaultseppunct}\relax
\EndOfBibitem
\bibitem[Heyd \latin{et~al.}(2003)Heyd, Scuseria, and
  Ernzerhof]{heyd2003JChemPhys}
Heyd,~J.; Scuseria,~G.~E.; Ernzerhof,~M. Hybrid functionals based on a screened
  Coulomb potential. \emph{The Journal of chemical physics} \textbf{2003},
  \emph{118}, 8207--8215\relax
\mciteBstWouldAddEndPuncttrue
\mciteSetBstMidEndSepPunct{\mcitedefaultmidpunct}
{\mcitedefaultendpunct}{\mcitedefaultseppunct}\relax
\EndOfBibitem
\bibitem[Tkatchenko and Scheffler(2009)Tkatchenko, and
  Scheffler]{tkatchenko2009PRL}
Tkatchenko,~A.; Scheffler,~M. Accurate molecular van der Waals interactions
  from ground-state electron density and free-atom reference data.
  \emph{Physical review letters} \textbf{2009}, \emph{102}, 073005\relax
\mciteBstWouldAddEndPuncttrue
\mciteSetBstMidEndSepPunct{\mcitedefaultmidpunct}
{\mcitedefaultendpunct}{\mcitedefaultseppunct}\relax
\EndOfBibitem
\bibitem[Tkatchenko \latin{et~al.}(2012)Tkatchenko, DiStasio~Jr, Car, and
  Scheffler]{tkatchenko2012PRL}
Tkatchenko,~A.; DiStasio~Jr,~R.~A.; Car,~R.; Scheffler,~M. Accurate and
  efficient method for many-body van der Waals interactions. \emph{Physical
  review letters} \textbf{2012}, \emph{108}, 236402\relax
\mciteBstWouldAddEndPuncttrue
\mciteSetBstMidEndSepPunct{\mcitedefaultmidpunct}
{\mcitedefaultendpunct}{\mcitedefaultseppunct}\relax
\EndOfBibitem
\bibitem[Weng and Gao(2018)Weng, and Gao]{weng2018honeycomb}
Weng,~J.; Gao,~S.-P. A honeycomb-like monolayer of HfO 2 and the calculation of
  static dielectric constant eliminating the effect of vacuum spacing.
  \emph{Physical Chemistry Chemical Physics} \textbf{2018}, \emph{20},
  26453--26462\relax
\mciteBstWouldAddEndPuncttrue
\mciteSetBstMidEndSepPunct{\mcitedefaultmidpunct}
{\mcitedefaultendpunct}{\mcitedefaultseppunct}\relax
\EndOfBibitem
\bibitem[Ren \latin{et~al.}(2020)Ren, Tang, Sun, Cai, Cheng, and
  Zhang]{ren2020direct}
Ren,~K.; Tang,~W.; Sun,~M.; Cai,~Y.; Cheng,~Y.; Zhang,~G. A direct Z-scheme PtS
  2/arsenene van der Waals heterostructure with high photocatalytic water
  splitting efficiency. \emph{Nanoscale} \textbf{2020}, \emph{12},
  17281--17289\relax
\mciteBstWouldAddEndPuncttrue
\mciteSetBstMidEndSepPunct{\mcitedefaultmidpunct}
{\mcitedefaultendpunct}{\mcitedefaultseppunct}\relax
\EndOfBibitem
\bibitem[Onida \latin{et~al.}(2002)Onida, Reining, and
  Rubio]{onida2002electronic}
Onida,~G.; Reining,~L.; Rubio,~A. Electronic excitations: density-functional
  versus many-body Green’s-function approaches. \emph{Reviews of modern
  physics} \textbf{2002}, \emph{74}, 601\relax
\mciteBstWouldAddEndPuncttrue
\mciteSetBstMidEndSepPunct{\mcitedefaultmidpunct}
{\mcitedefaultendpunct}{\mcitedefaultseppunct}\relax
\EndOfBibitem
\bibitem[Jiang \latin{et~al.}(2012)Jiang, Rinke, and
  Scheffler]{jiang2012electronic}
Jiang,~H.; Rinke,~P.; Scheffler,~M. Electronic properties of lanthanide oxides
  from the G W perspective. \emph{Physical Review B} \textbf{2012}, \emph{86},
  125115\relax
\mciteBstWouldAddEndPuncttrue
\mciteSetBstMidEndSepPunct{\mcitedefaultmidpunct}
{\mcitedefaultendpunct}{\mcitedefaultseppunct}\relax
\EndOfBibitem
\bibitem[Rahman \latin{et~al.}(2018)Rahman, Morbec, Rahman, and
  Kratzer]{rahman2018commensurate}
Rahman,~A.~U.; Morbec,~J.~M.; Rahman,~G.; Kratzer,~P. Commensurate versus
  incommensurate heterostructures of group-III monochalcogenides.
  \emph{Physical Review Materials} \textbf{2018}, \emph{2}, 094002\relax
\mciteBstWouldAddEndPuncttrue
\mciteSetBstMidEndSepPunct{\mcitedefaultmidpunct}
{\mcitedefaultendpunct}{\mcitedefaultseppunct}\relax
\EndOfBibitem
\bibitem[Bastos \latin{et~al.}(2019)Bastos, Besse, Da~Silva, and
  Sipahi]{bastos2019ab}
Bastos,~C.~M.; Besse,~R.; Da~Silva,~J.~L.; Sipahi,~G.~M. Ab initio
  investigation of structural stability and exfoliation energies in transition
  metal dichalcogenides based on Ti-, V-, and Mo-group elements. \emph{Physical
  Review Materials} \textbf{2019}, \emph{3}, 044002\relax
\mciteBstWouldAddEndPuncttrue
\mciteSetBstMidEndSepPunct{\mcitedefaultmidpunct}
{\mcitedefaultendpunct}{\mcitedefaultseppunct}\relax
\EndOfBibitem
\bibitem[Haastrup \latin{et~al.}(2018)Haastrup, Strange, Pandey, Deilmann,
  Schmidt, Hinsche, Gjerding, Torelli, Larsen, Riis-Jensen, \latin{et~al.}
  others]{haastrup2018computational}
Haastrup,~S.; Strange,~M.; Pandey,~M.; Deilmann,~T.; Schmidt,~P.~S.;
  Hinsche,~N.~F.; Gjerding,~M.~N.; Torelli,~D.; Larsen,~P.~M.;
  Riis-Jensen,~A.~C., \latin{et~al.}  The Computational 2D Materials Database:
  high-throughput modeling and discovery of atomically thin crystals. \emph{2D
  Materials} \textbf{2018}, \emph{5}, 042002\relax
\mciteBstWouldAddEndPuncttrue
\mciteSetBstMidEndSepPunct{\mcitedefaultmidpunct}
{\mcitedefaultendpunct}{\mcitedefaultseppunct}\relax
\EndOfBibitem
\bibitem[Xia \latin{et~al.}(2018)Xia, Xiong, Xiao, Du, Fang, Li, and
  Jia]{xia2018enhanced}
Xia,~C.; Xiong,~W.; Xiao,~W.; Du,~J.; Fang,~L.; Li,~J.; Jia,~Y. Enhanced
  carrier concentration and electronic transport by inserting graphene into van
  der Waals heterostructures of transition-metal dichalcogenides.
  \emph{Physical Review Applied} \textbf{2018}, \emph{10}, 024028\relax
\mciteBstWouldAddEndPuncttrue
\mciteSetBstMidEndSepPunct{\mcitedefaultmidpunct}
{\mcitedefaultendpunct}{\mcitedefaultseppunct}\relax
\EndOfBibitem
\bibitem[Hu \latin{et~al.}(2016)Hu, Kou, and Sun]{hu2016stacking}
Hu,~X.; Kou,~L.; Sun,~L. Stacking orders induced direct band gap in bilayer
  MoSe 2-WSe 2 lateral heterostructures. \emph{Scientific reports}
  \textbf{2016}, \emph{6}, 31122\relax
\mciteBstWouldAddEndPuncttrue
\mciteSetBstMidEndSepPunct{\mcitedefaultmidpunct}
{\mcitedefaultendpunct}{\mcitedefaultseppunct}\relax
\EndOfBibitem
\bibitem[Padilha \latin{et~al.}(2015)Padilha, Fazzio, and
  da~Silva]{padilha2015van}
Padilha,~J.; Fazzio,~A.; da~Silva,~A.~J. van der Waals heterostructure of
  phosphorene and graphene: tuning the Schottky barrier and doping by
  electrostatic gating. \emph{Physical review letters} \textbf{2015},
  \emph{114}, 066803\relax
\mciteBstWouldAddEndPuncttrue
\mciteSetBstMidEndSepPunct{\mcitedefaultmidpunct}
{\mcitedefaultendpunct}{\mcitedefaultseppunct}\relax
\EndOfBibitem
\bibitem[Ganose \latin{et~al.}(2018)Ganose, Jackson, and
  Scanlon]{ganose2018sumo}
Ganose,~A.; Jackson,~A.; Scanlon,~D. sumo: Command-line tools for plotting and
  analysis of periodic* ab initio* calculations. \emph{Journal of Open Source
  Software} \textbf{2018}, \emph{3}, 717\relax
\mciteBstWouldAddEndPuncttrue
\mciteSetBstMidEndSepPunct{\mcitedefaultmidpunct}
{\mcitedefaultendpunct}{\mcitedefaultseppunct}\relax
\EndOfBibitem
\bibitem[Lu \latin{et~al.}(2017)Lu, Houssa, Luisier, and
  Pourtois]{lu2017impact}
Lu,~A. K.~A.; Houssa,~M.; Luisier,~M.; Pourtois,~G. Impact of Layer Alignment
  on the Behavior of MoS 2- ZrS 2 Tunnel Field-Effect Transistors: An Ab Initio
  Study. \emph{Physical Review Applied} \textbf{2017}, \emph{8}, 034017\relax
\mciteBstWouldAddEndPuncttrue
\mciteSetBstMidEndSepPunct{\mcitedefaultmidpunct}
{\mcitedefaultendpunct}{\mcitedefaultseppunct}\relax
\EndOfBibitem
\bibitem[Zhang \latin{et~al.}(2012)Zhang, Liu, and Zhou]{zhang2012D}
Zhang,~H.; Liu,~L.; Zhou,~Z. First-principles studies on facet-dependent
  photocatalytic properties of bismuth oxyhalides (BiOXs). \emph{RSC advances}
  \textbf{2012}, \emph{2}, 9224--9229\relax
\mciteBstWouldAddEndPuncttrue
\mciteSetBstMidEndSepPunct{\mcitedefaultmidpunct}
{\mcitedefaultendpunct}{\mcitedefaultseppunct}\relax
\EndOfBibitem
\bibitem[Dai and Zeng(2015)Dai, and Zeng]{dai2015titanium}
Dai,~J.; Zeng,~X.~C. Titanium trisulfide monolayer: theoretical prediction of a
  new direct-gap semiconductor with high and anisotropic carrier mobility.
  \emph{Angewandte Chemie} \textbf{2015}, \emph{127}, 7682--7686\relax
\mciteBstWouldAddEndPuncttrue
\mciteSetBstMidEndSepPunct{\mcitedefaultmidpunct}
{\mcitedefaultendpunct}{\mcitedefaultseppunct}\relax
\EndOfBibitem
\bibitem[Basera \latin{et~al.}(2019)Basera, Saini, and
  Bhattacharya]{basera2019self}
Basera,~P.; Saini,~S.; Bhattacharya,~S. Self energy and excitonic effect in
  (un) doped TiO 2 anatase: a comparative study of hybrid DFT, GW and BSE to
  explore optical properties. \emph{Journal of Materials Chemistry C}
  \textbf{2019}, \emph{7}, 14284--14293\relax
\mciteBstWouldAddEndPuncttrue
\mciteSetBstMidEndSepPunct{\mcitedefaultmidpunct}
{\mcitedefaultendpunct}{\mcitedefaultseppunct}\relax
\EndOfBibitem
\bibitem[Khanchandani \latin{et~al.}(2014)Khanchandani, Srivastava, Kumar,
  Ghosh, and Ganguli]{khanchandani2014band}
Khanchandani,~S.; Srivastava,~P.~K.; Kumar,~S.; Ghosh,~S.; Ganguli,~A.~K. Band
  gap engineering of ZnO using core/shell morphology with environmentally
  benign Ag2S sensitizer for efficient light harvesting and enhanced
  visible-light photocatalysis. \emph{Inorganic Chemistry} \textbf{2014},
  \emph{53}, 8902--8912\relax
\mciteBstWouldAddEndPuncttrue
\mciteSetBstMidEndSepPunct{\mcitedefaultmidpunct}
{\mcitedefaultendpunct}{\mcitedefaultseppunct}\relax
\EndOfBibitem
\end{mcitethebibliography}
\newpage


\end{document}









\begin{center}
	{\Large \bf Supporting Information}\\ 
\end{center}
\vspace*{12pt}

\noindent {\bf \Large I. Electrostatic Potential\\}
\noindent {\bf \Large II. Lattice Parameters\\}
\noindent {\bf \Large III. Band Gaps of Configurations\\}
\noindent {\bf \Large IV. Bandstructures of Z-scheme vdW HTSs\\}
\noindent {\bf \Large V. Planar Averaged Charged Density\\}
\noindent {\bf \Large VI. Absorption Spectra of Monolayers\\}
\noindent {\bf \Large VII. Exciton Binding Energy\\}
\noindent {\bf \Large VIII. Carrier Mobility\\}
\newpage
\begin{center}
	\section{\bf \Large I. Electrostatic Potential}
\end{center}

\vspace*{2cm}
\begin{figure}[!htp]
	\begin{center} 
		
		\includegraphics[scale=0.60]{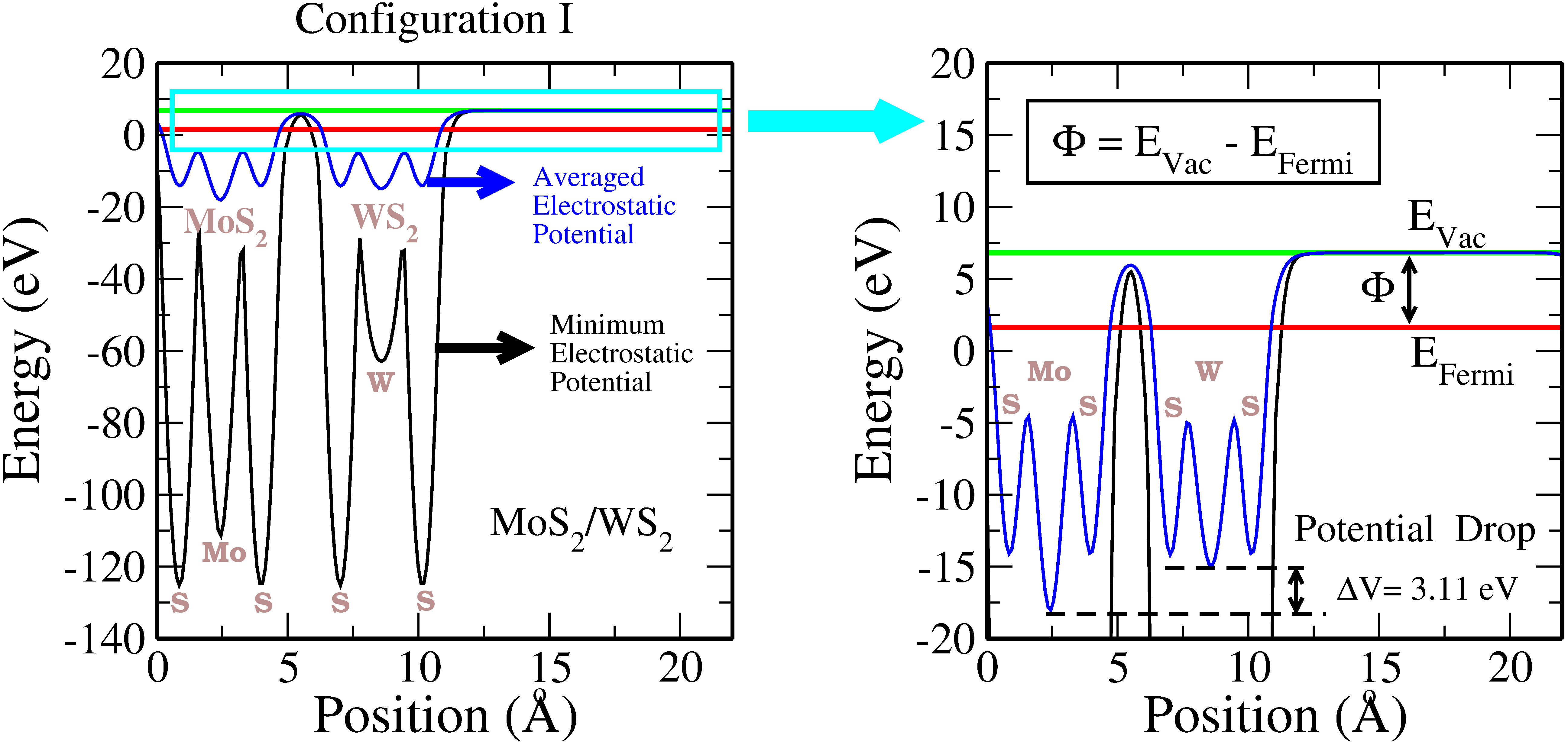}
		\caption{\textrm{Schematic of electrostatic potential as obtained corresponding to MoS{$_2$}/WS{$_2$}  vdW HTS.}}
		\label{S1}
	\end{center} 
\end{figure}
\clearpage
\begin{center}
	\noindent {\bf \Large Averaged Electrostatic Potential}
\end{center}
\begin{figure}[!htp]
	\begin{center} 
		
		\includegraphics[scale=0.60]{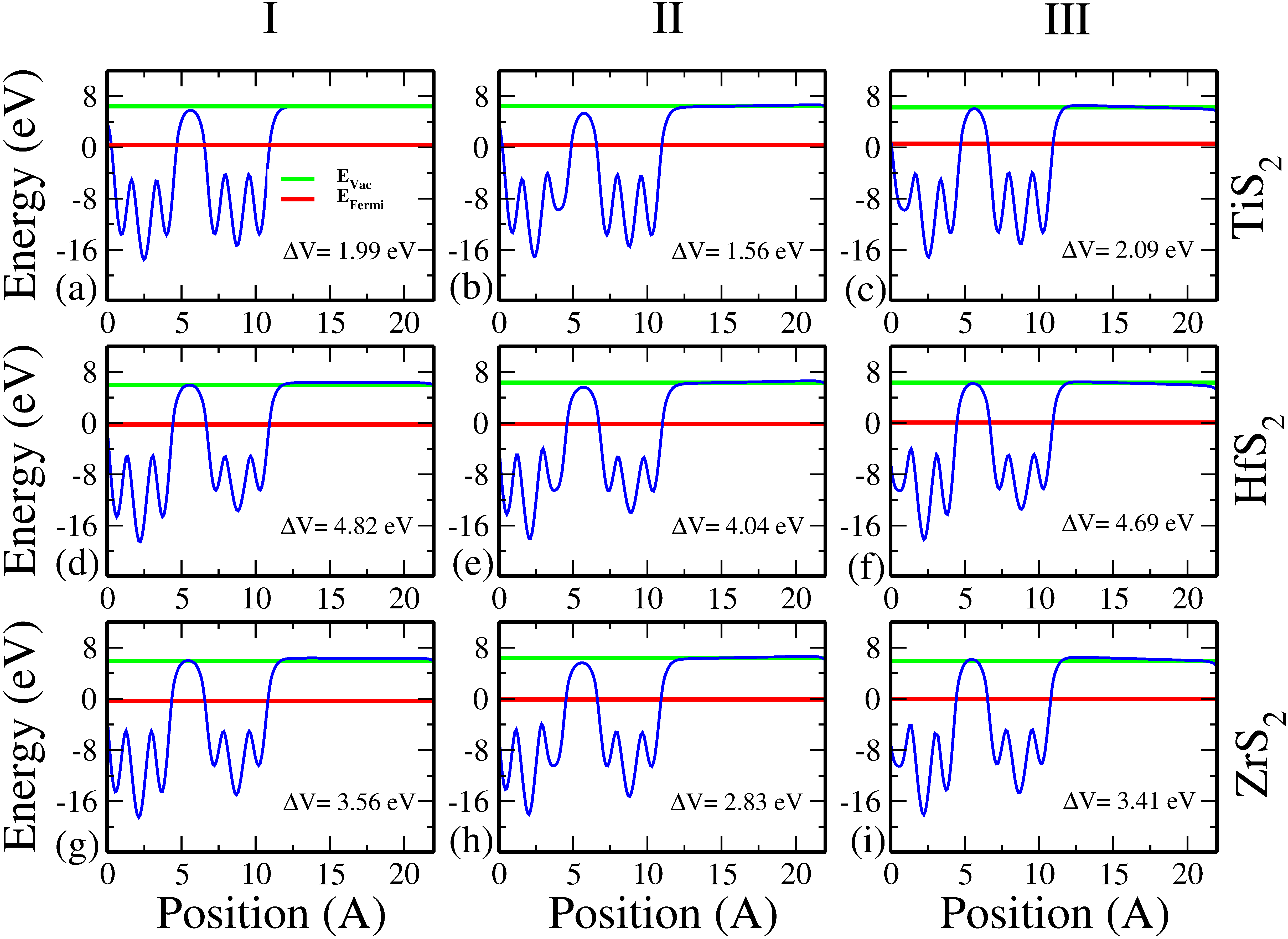}
		\caption{\textrm{Averaged electrostatic potential corresponding to TMDs viz. TiS{$_2$} (upper panel), HfS{$_2$} (middle panel) and ZrS{$_2$} (lower panel) for I, II and III configurations.}}
		\label{S2}
	\end{center} 
\end{figure}
\begin{figure}[!htp]
	\begin{center} 
		
		\includegraphics[scale=0.60]{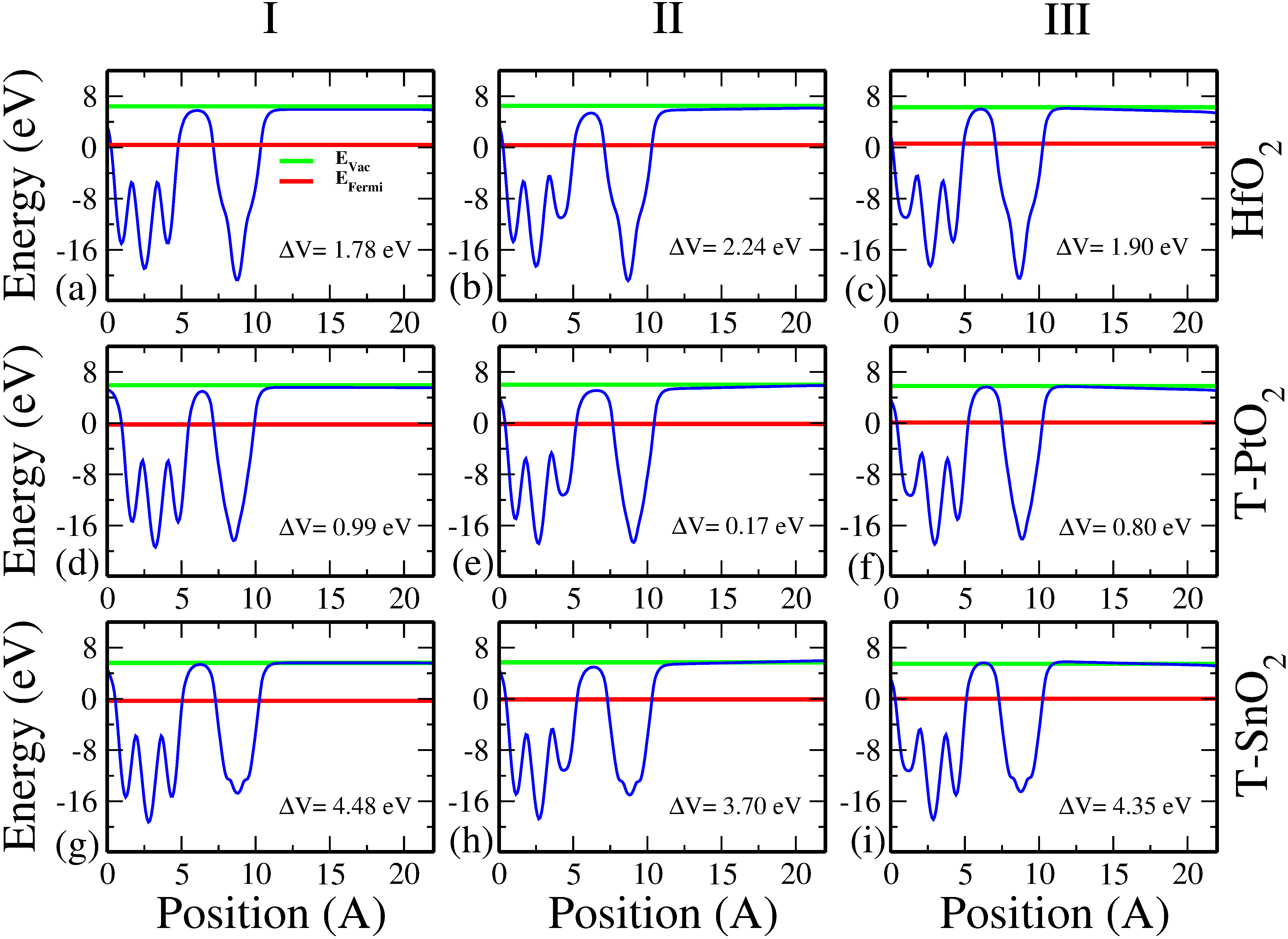}
		\caption{\textrm{Averaged electrostatic potential corresponding to TMOs viz. HfO{$_2$} (upper panel), T-SnO{$_2$} (middle panel) and T-PtO{$_2$} (lower panel) for I, II and III configurations.}}
		\label{S3}
	\end{center} 
\end{figure}
\clearpage
\begin{center}
	\noindent {\bf \Large Minimum Electrostatic Potential}
\end{center}
\begin{figure}[!htp]
	\begin{center} 
		\includegraphics[scale=0.40]{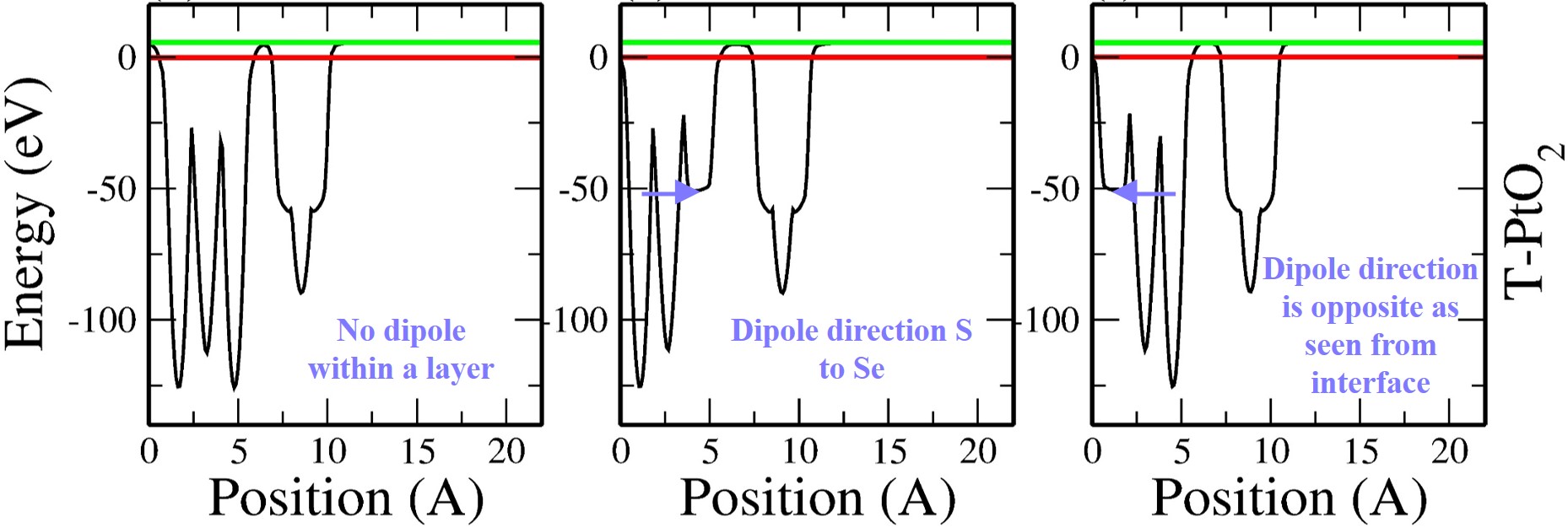}
		\caption{\textrm{Electrostatic potential of T-PtO{$_2$} for I, II and III configurations, showing dipole direction at the interface.}}
		\label{S4}
	\end{center} 
\end{figure}
\begin{figure}[!htp]
	\begin{center} 
		
		\includegraphics[scale=0.49]{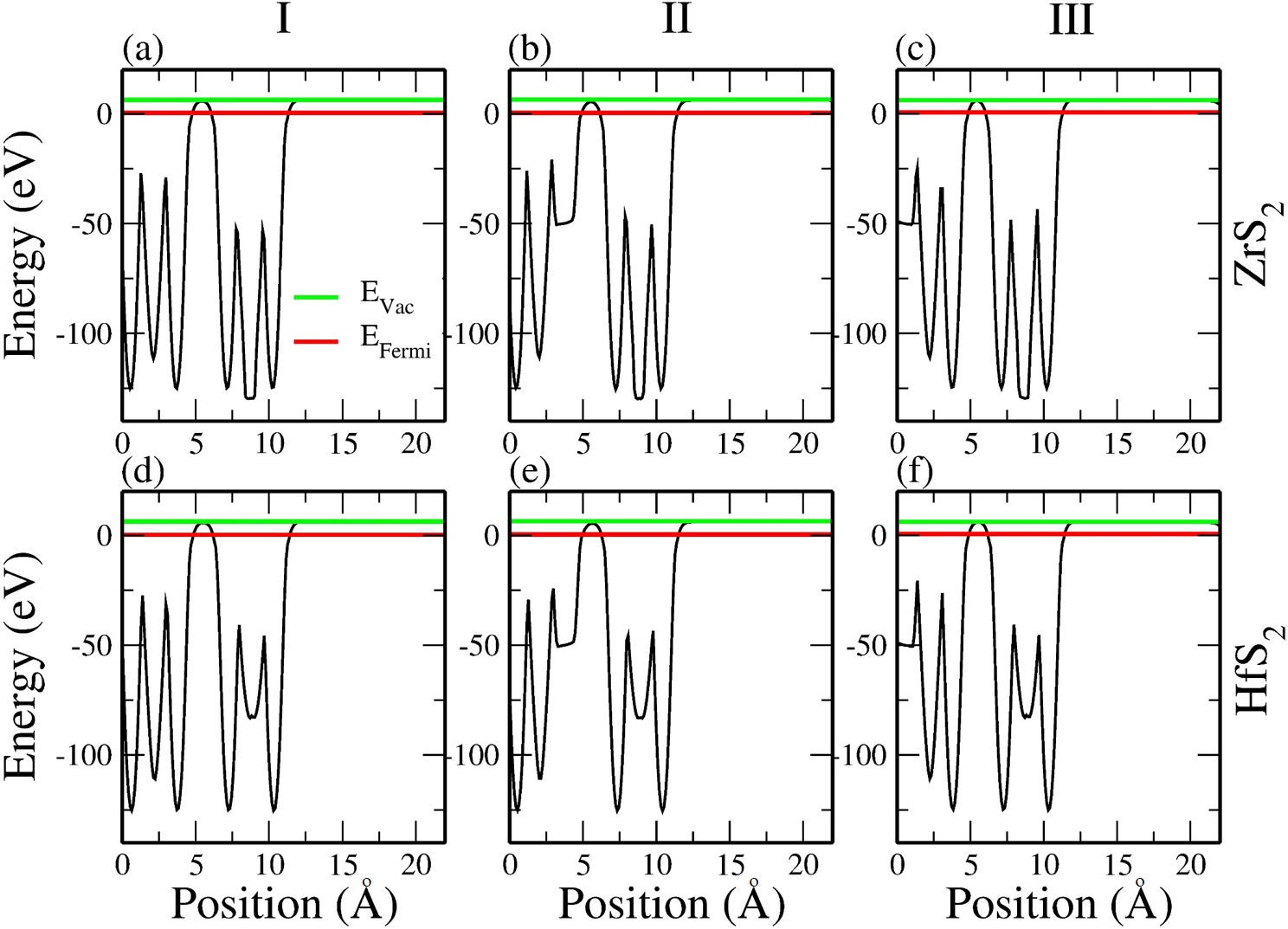}
		\caption{\textrm{Electrostatic potential corresponding to ZrS{$_2$} (upper panel) and HfS{$_2$} (lower panel) for I, II and III configurations.}}
		\label{S5}
	\end{center} 
\end{figure}
\begin{figure}[!htp]
	\begin{center} 
		
		\includegraphics[scale=0.65]{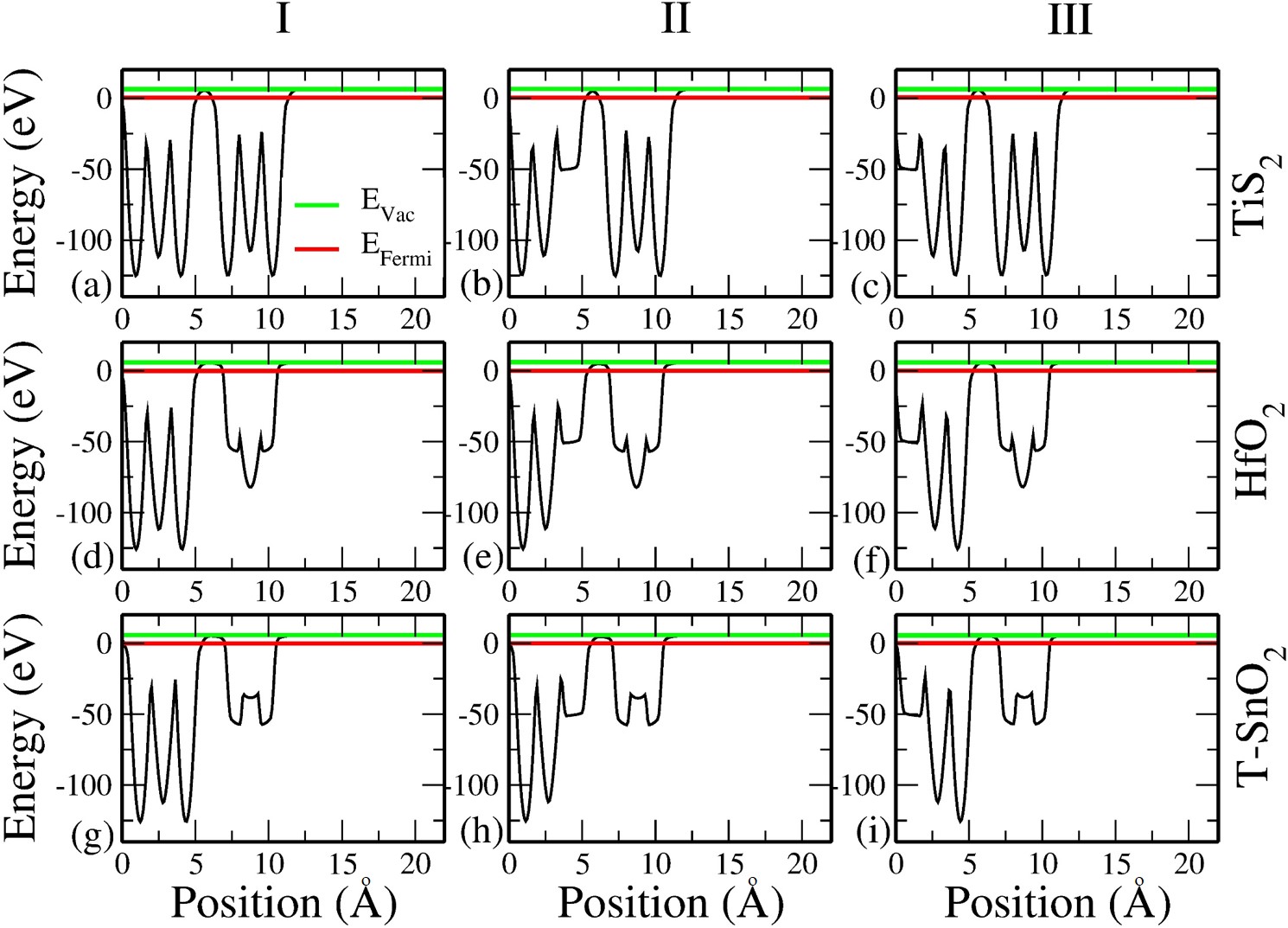}
		\caption{\textrm{Electrostatic potential corresponding to TiS{$_2$} (upper panel), HfO{$_2$} (middle panel) and SnO{$_2$} (lower panel) for I, II and III configurations.}}
		\label{S6}
	\end{center} 
\end{figure}
\clearpage
\newpage
\begin{center}
	\section{\bf \Large  II. Lattice Parameters}
\end{center}

\begin{table}[htbp]
	\caption{Lattice constants (l) of 2$\times$2 monolayers} 
	\begin{center}
		\begin{tabular}[c]{p{0.08\textwidth}p{0.08\textwidth}p{0.08\textwidth}p{0.08\textwidth}p{0.08\textwidth}p{0.08\textwidth}p{0.08\textwidth}p{0.09\textwidth}p{0.09\textwidth}}\hline 
			\\[-1em]
			BX{$_2$}&MoS{$_2$}&WS{$_2$}&ZrS{$_2$}&HfS{$_2$}&TiS{$_2$}&HfO{$_2$}&T-PtO{$_2$}&T-SnO{$_2$} \\ \hline
			\\[-1em]
			l&6.321&6.321&7.120&7.019&6.660&6.243&6.296&6.450 \\  \hline
		\end{tabular}
		\label{Table5}
	\end{center}
\end{table}
\begin{center}
	\section{\bf \Large  III. Band Gaps of Configurations}
\end{center}

\begin{table}[htbp]
	\caption{Band gaps of the monolayers and their corresponding vdW HTSs.} 
	\begin{center}
		\begin{tabular}[c]{|p{0.08\textwidth}|p{0.13\textwidth}|p{0.13\textwidth}|p{0.13\textwidth}|p{0.13\textwidth}|}\hline 
			\\[-1em]
			\multirow{2}{*}{BX{$_2$}}&\multicolumn{4}{c|}{Band Gap (eV) (Indirect/Direct)}\\
			&  Monolayer  &   I & II & III \\ \hline
			\\[-1em]
			MoS{$_2$} &   -/2.257  &   - &-&- \\ \hline
			\\[-1em]
			MoSSe  	  &   -/2.172 &   - &-&- \\ \hline
			\\[-1em]
			WS{$_2$}  &   -/2.447  &   1.636/2.038  & 1.879/1.885 & 1.589/1.928\\ \hline
			\\[-1em]
			ZrS{$_2$} &   1.829/1.892  &   0.752/0.809  & 0.123/0.177 &0.676/0.729\\ \hline
			\\[-1em]
			HfS{$_2$} &   1.937/2.019  &   0.746/0.863  & 0.079/0.185 &0.663/0.773\\ \hline
			\\[-1em]
			TiS{$_2$} &   1.672/1.769  &   0.414/0.418  & 0.047/0.056&0.520/0.525\\ \hline
			\\[-1em]
			HfO{$_2$} &   3.516/3.520 &   0.486/0.499  & 0.072/0.090 &0.440/0.453\\ \hline
			\\[-1em]
			T-PtO{$_2$} & 3.304/3.313  &   1.323/1.525  & 0.595/0.846 & 1.178/1.423\\ \hline
			\\[-1em]
			T-SnO{$_2$} & 4.143/4.297  &   1.130/1.285  & 0.460/0.882 &  1.069/1.473\\ \hline
			
		\end{tabular}
		\label{Table6}
	\end{center}
\end{table}
\clearpage
\newpage
\begin{center}
	\section{\bf \Large  IV. Bandstructures of  Z-scheme vdW HTSs}
\end{center}

\begin{figure}[!htp]
	\includegraphics[scale=0.50]{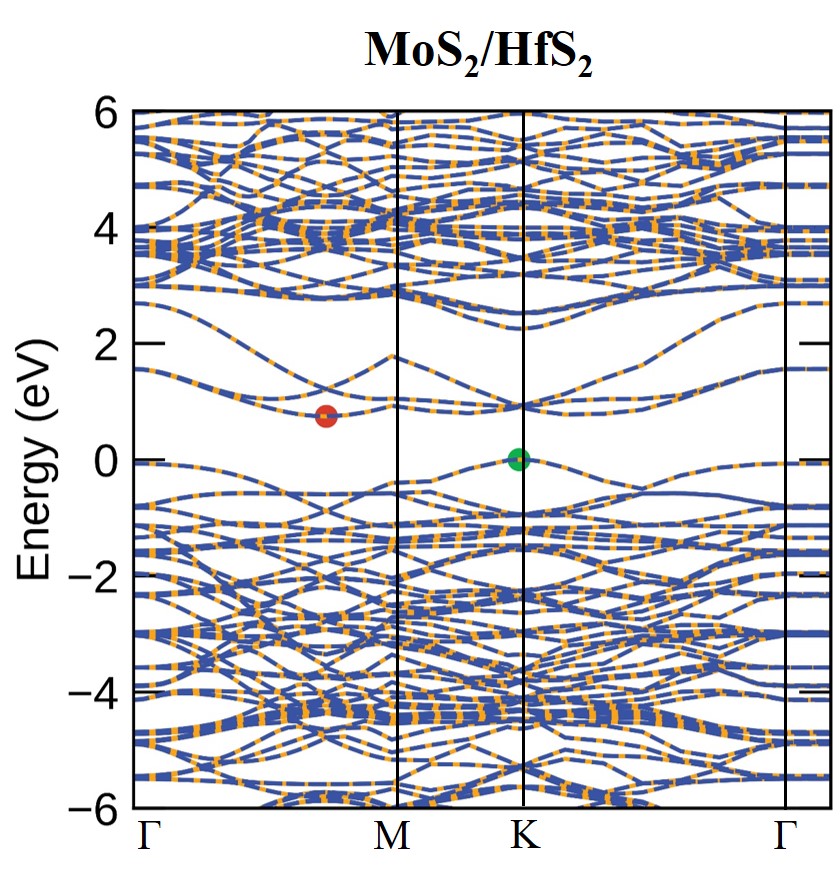}
	\hfill
	\qquad
	\includegraphics[scale=0.50]{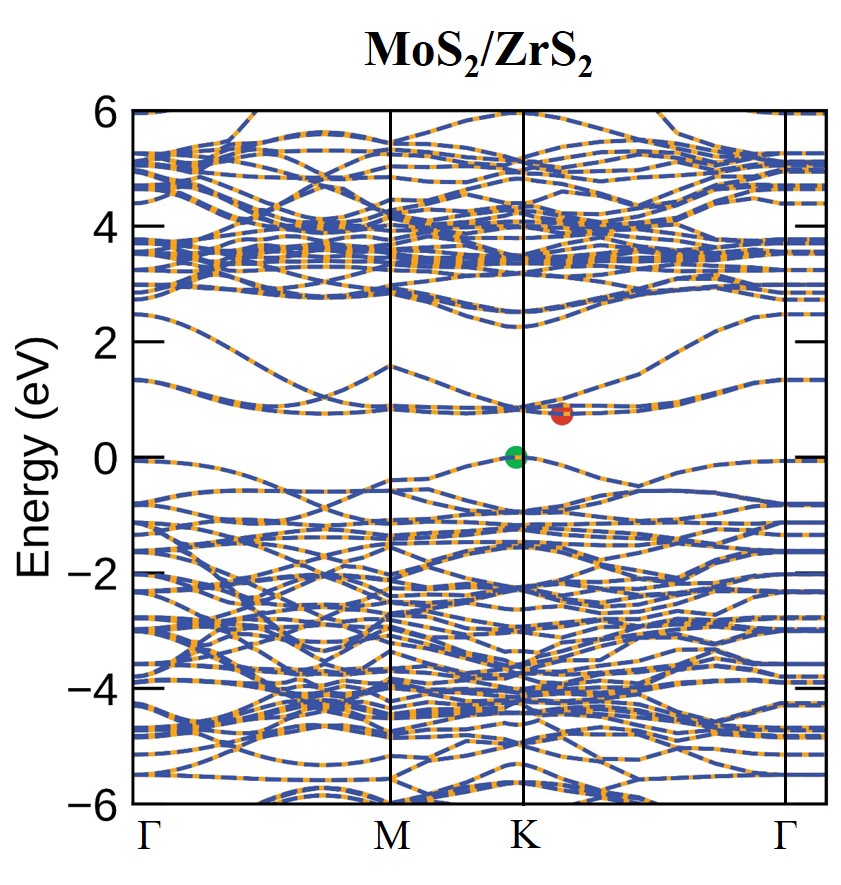}
	\hfill
	\qquad
	\includegraphics[scale=0.50]{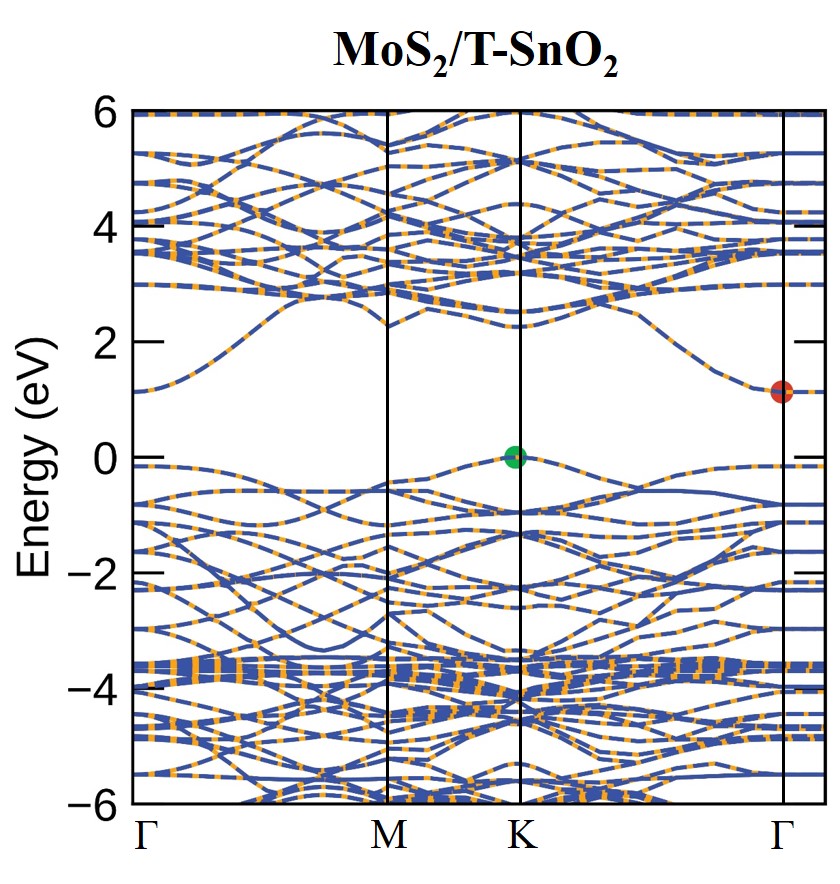}
	\hfill
	\qquad
	\includegraphics[scale=0.50]{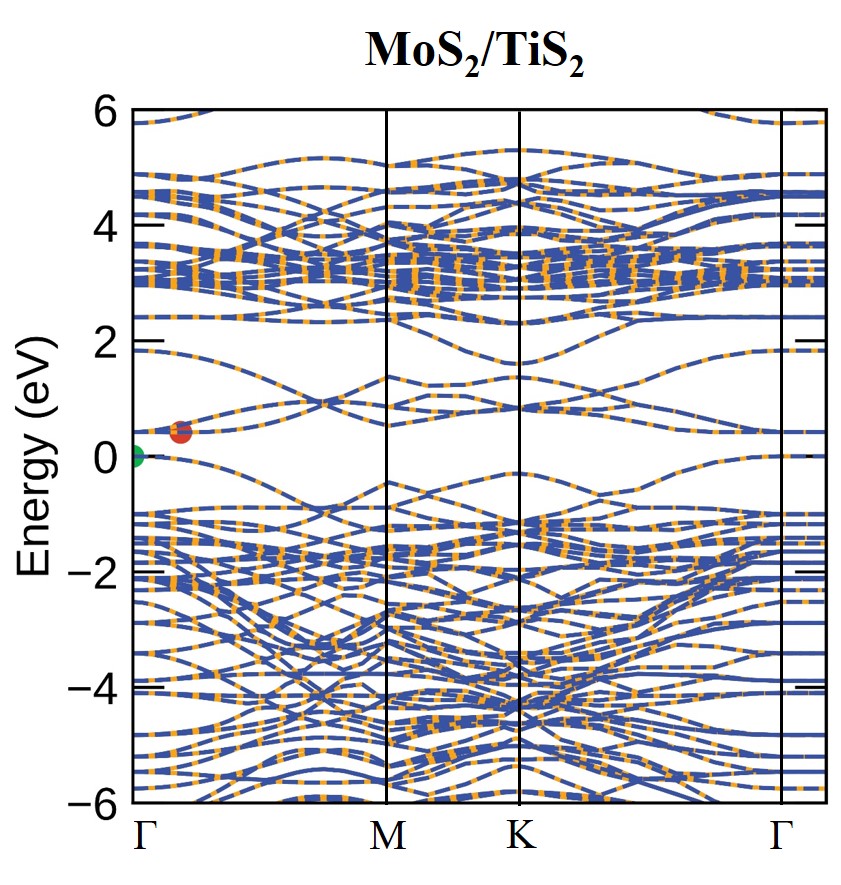}
	\hfill
	\caption[]{\textrm{(Color online) Bandstructures  corresponding to the supercell of MoS{$_2$}/BX{$_2$} HTSs (configuration I) where MoS{$_2$}/ZrS{$_2$} has similar bandstructure as that of MoS{$_2$}/HfS{$_2$}. The bandstructures corresponding to the MoSSe based vdW HTSs (configurations II and III) are similar with slight change in energetics. The red and green points correspond to the conduction band minimum and valence band maximum.}}
	\label{fig:1}
\end{figure}
\clearpage
\newpage
\begin{center}
	\section{\bf \Large  V. Planar Averaged Charged Density}
\end{center}
\vspace{1cm}
\begin{figure}[!htp]
		\includegraphics[scale=0.29]{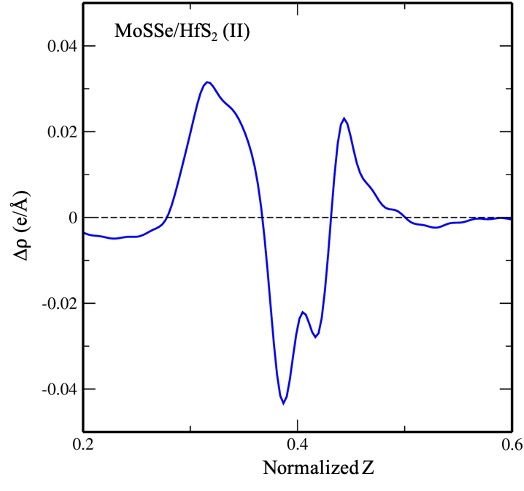}
		\includegraphics[scale=0.29]{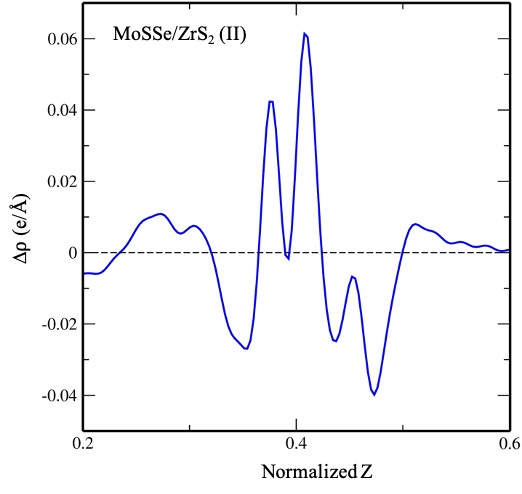}
		\includegraphics[scale=0.29]{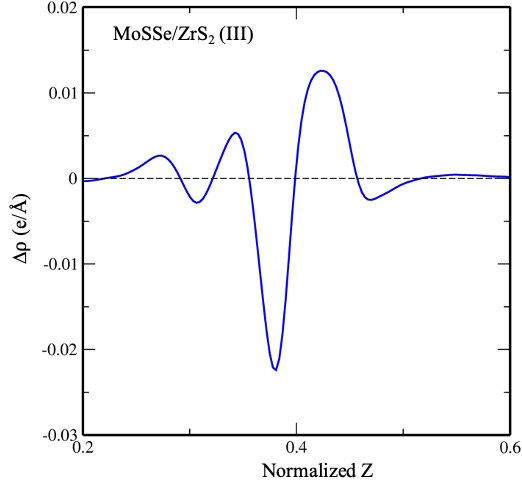}
		\includegraphics[scale=0.29]{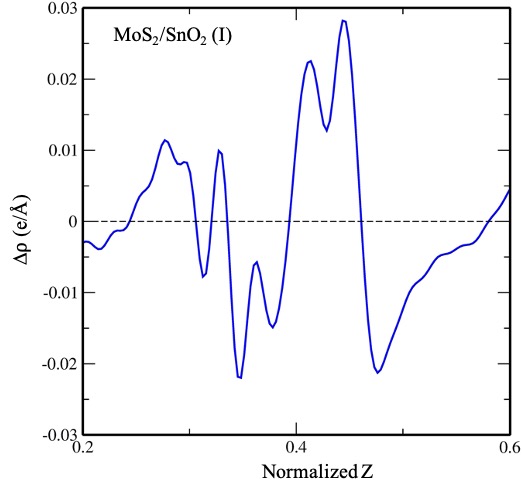}
		\includegraphics[scale=0.29]{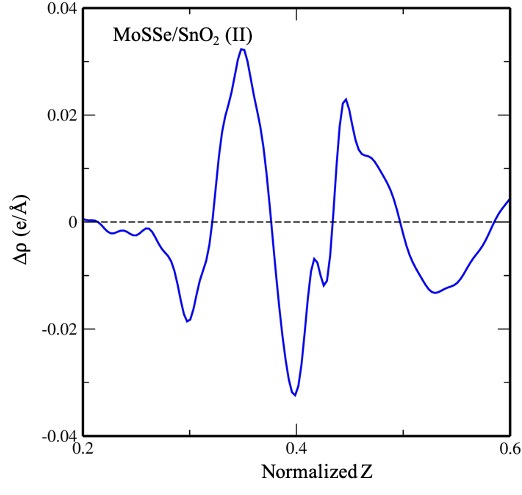}
		\includegraphics[scale=0.29]{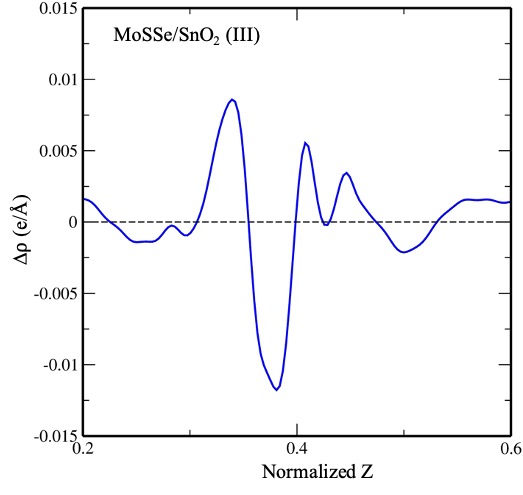}
		\includegraphics[scale=0.29]{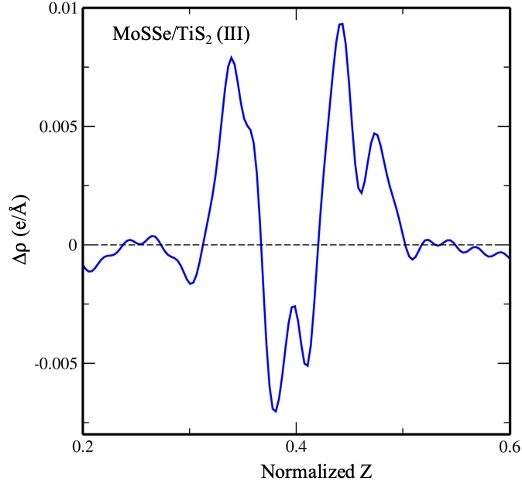}
		\caption[]{\textrm{(Color online) Planar averaged charge density}}
		\label{fig:chd}
\end{figure}
\clearpage
\newpage
\begin{center}
	\section{\bf \Large  VI. Absorption Spectra of Monolayers}
\end{center}
\vspace{1cm}
\begin{figure}[!htp]
	\begin{center} 
		
		\includegraphics[scale=0.65]{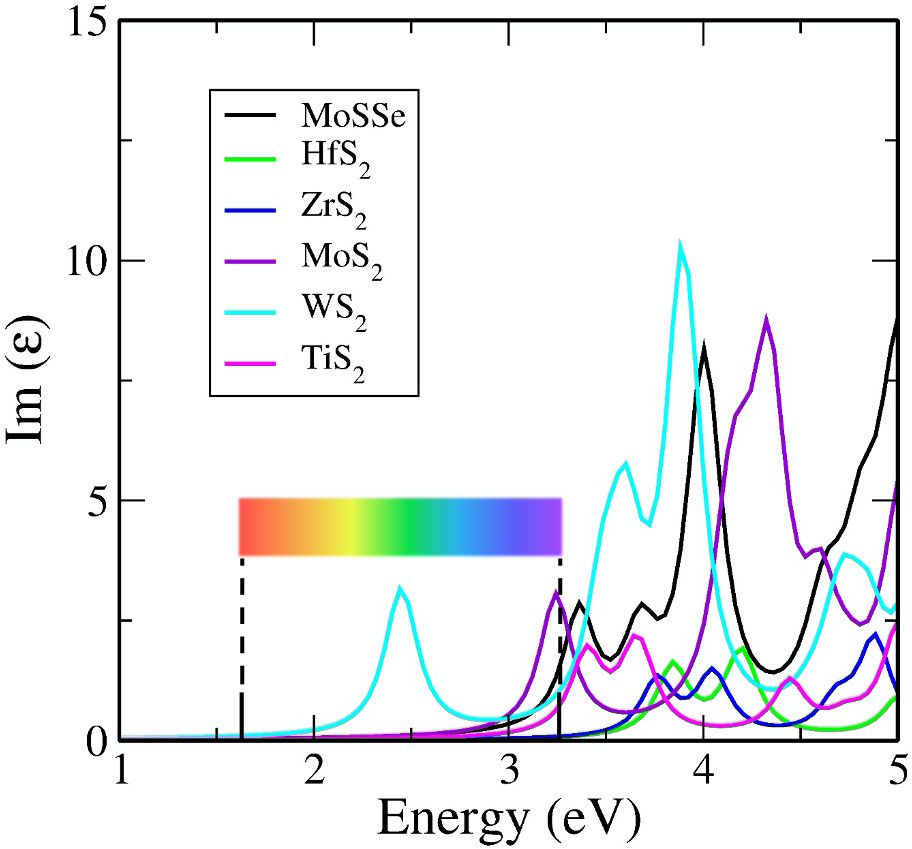}
		\caption{\textrm{Absorption spectra of Janus (MoSSe) and monolayer TMDs i.e MoS$_2$, WS$_2$, TiS$_2$, HfS$_2$ and ZrS$_2$}}
		\label{S8}
	\end{center} 
\end{figure}
\clearpage
\begin{center}
	\section{\bf \Large  VII. Exciton Binding Energy}
\end{center}
\noindent We have performed mBSE calculations, with an intention to simply compare the exciton binding energy (E{$_\textrm{B}$}) of vdW HTSs with MoS$_2$ and MoSSe. We have included four valence and conduction bands for mBSE on top of hybrid functional. However, there is a huge scope in understanding the excitonic transitions in these systems with different computational approaches. Presently, considering the Z-scheme application, the smaller E{$_\textrm{B}$} of vdW HTSs as compared to MoS{$_2$} (or MoSSe) would allow more e{$^-$} - h{$^+$} recombination in vdW HTS thereby facilitating MoS{$_2$} (or MoSSe) for HER.
\\
\begin{table}[htbp]
	\caption{Exciton binding energies of vdW HTSs, MoS{$_2$} and MoSSe monolayers} 
	\begin{center}
		\begin{tabular}[c]{p{0.03\textwidth}p{0.04\textwidth}p{0.06\textwidth}p{0.12\textwidth}p{0.12\textwidth}p{0.12\textwidth}p{0.12\textwidth}p{0.12\textwidth}p{0.12\textwidth}}\hline 
			\\[-1em]
			&MoS{$_2$}&MoSSe&MoSSe/HfS{$_2$} (II)&MoSSe/TiS{$_2$} (III)&MoS{$_2$}/SnO{$_2$} (I)&MoSSe/ZrS{$_2$} (II)&MoSSe/ZrS{$_2$} (III)&MoSSe/SnO{$_2$} (III) \\ \hline
			\\[-1em]
			E{$_\textrm{B}$} (eV)&1.5&1.8&0.44&0.37&1.3&0.42&0.41&1.3\\  \hline
		\end{tabular}
		\label{Ebe}
	\end{center}
\end{table}

\clearpage
\newpage
\begin{center}
	\section{\bf \Large  VIII. Carrier Mobility}
\end{center}

\begin{myequation}
	\mu = \frac{2\textrm{e}\hbar{^3}C}{3k{_B}T|m^*|^2E_1^2} 
\end{myequation}\\
\vspace{0.5cm}
\\
\noindent In this expression C is defined as C = [$\partial^2$E/$\partial$$\delta$$^2$ ]/S{$^0$}. E refers to the total energy of the system, $\delta$ is the applied uniaxial strain, and S{$^0$} is the area of the optimized vdW HTS. m* is the effective mass, expressed as m* = $\hbar^2$ ($\partial^2$E/$\partial$k$^2$){$^{-1}$}. T represents temperature, and E{$_1$} is the deformation potential constant that is defined as {$\Delta$}E = E{$_1$}({$\Delta$}l/l{$_0$} ). Here, {$\Delta$}E is the energy shift of the band edge position with respect to the lattice strain {$\Delta$}l/l{$_0$}. The energies of the band edges (CBm or VBM) are obtained with vacuum level as the reference.